\pdfoutput=1

\documentclass[a4paper,fleqn,usenatbib]{mn2e}
\usepackage{graphicx}
\usepackage{float}
\usepackage{subfigure}
\usepackage{amssymb}
\usepackage{amsmath}
\usepackage{afterpage}
\usepackage[usenames]{color}
\usepackage[totalwidth=485pt, totalheight=680pt]{geometry}
\usepackage{yfonts}
\usepackage{mathrsfs}
\usepackage{upgreek}
\usepackage{lastpage}

\usepackage{hyperref}
\usepackage[all]{hypcap}

\newcommand\fauxsc[1]{\fauxschelper#1 \relax\relax}
\def\fauxschelper#1 #2\relax{%
  \fauxschelphelp#1\relax\relax%
  \if\relax#2\relax\else\ \fauxschelper#2\relax\fi%
}
\def\Hscale{.85}\def\Vscale{.74}\def\Cscale{1.12}
\def\fauxschelphelp#1#2\relax{%
  \ifnum`#1>``\ifnum`#1<`\{\scalebox{\Hscale}[\Vscale]{\uppercase{#1}}\else%
    \scalebox{\Cscale}[1]{#1}\fi\else\scalebox{\Cscale}[1]{#1}\fi%
  \ifx\relax#2\relax\else\fauxschelphelp#2\relax\fi}

\title{A statistical investigation of the mass discrepancy--acceleration relation}
\author[H.~Desmond]{Harry Desmond\thanks{E-mail: harryd2@stanford.edu}\\
Kavli Institute for Particle Astrophysics and Cosmology and Physics Department, Stanford University, Stanford, CA 94305, USA; \\
SLAC National Accelerator Laboratory, Menlo Park, CA 94025, USA
}

\pubyear{2016}

\begin{document}
\label{FirstPage}
\pagerange{\pageref{FirstPage}--\pageref{LastPage}}
\maketitle

\begin{abstract}
We use the mass discrepancy--acceleration relation (the correlation between the ratio of total-to-visible mass and acceleration in galaxies; MDAR) to test the galaxy--halo connection. We analyse the MDAR using a set of 16 statistics that quantify its four most important features: its shape, its scatter, the presence of a ``characteristic acceleration scale,'' and the correlation of its residuals with other galaxy properties. We construct an empirical framework for the galaxy--halo connection in $\Lambda$CDM to generate predictions for these statistics, starting with conventional correlations (halo abundance matching; AM) and introducing more where required. Comparing to the SPARC data, we find that: 1) the approximate shape of the MDAR is readily reproduced by AM, and there is no evidence that the acceleration at which dark matter becomes negligible has less spread in the data than in AM mocks; 2) even under conservative assumptions, AM significantly overpredicts the scatter in the relation and its normalisation at low acceleration, and furthermore positions dark matter too close to galaxies' centres on average; 3) the MDAR affords $2\sigma$ evidence for an anticorrelation of galaxy size and Hubble type with halo mass or concentration at fixed stellar mass. Our analysis lays the groundwork for a bottom-up determination of the galaxy--halo connection from relations such as the MDAR, provides concrete statistical tests for specific galaxy formation models, and brings into sharper focus the relative evidence accorded by galaxy kinematics to $\Lambda$CDM and modified gravity alternatives.
\end{abstract}


\section{INTRODUCTION}
\label{sec:intro}

The internal motions of galaxies are largely set by dark matter, which outweighs baryonic matter by at least five to one. A key goal of galaxy astrophysics is to relate the visible and dark mass in any given system, which to first order means determining the correlations between the structural parameters of galaxies (e.g. $M_*$, $M_\mathrm{gas}$, $R_\mathrm{d}$ and Hubble type $T$) and those of dark matter haloes (e.g. $M_\mathrm{vir}$, $c$, and $\lambda$). This programme requires extensive observation of the rotation and velocity dispersion profiles of galaxies and their baryonic mass distributions, combined with detailed dynamical modelling under a variety of assumptions about the galaxy--halo connection. 

Traditional summaries of the relation between the baryonic mass distribution and internal motion of galaxies relate one-point statistics of these functions, reducing the former to a total galaxy mass and size, and the latter to a single measure of velocity. These are the Tully--Fisher, mass--size and Faber--Jackson relations, and Fundamental Plane. More information, however, can be found in the full \emph{radial} interdependence of mass and velocity, and analysis of this may be expected to afford not only a more stringent test of specific galaxy formation models, but also a richer foundation for a bottom-up determination of the galaxy--halo connection.

The local correlation of dark and visible matter is usefully described by the ratio of enclosed dynamical mass (determined kinematically) to enclosed baryonic mass (determined photometrically). A proxy for this quantity is $V^2_\mathrm{tot}(r)/V^2_\mathrm{bar}(r)$, to which it is equal in the case of spherical mass distribution. $V^2_\mathrm{tot}(r)/V^2_\mathrm{bar}(r)$ is known as the ``mass discrepancy''~\citep{MG99}, and will be denoted by $\mathcal{D}$ hereafter. In general, $\mathcal{D}$ may be any function of position $r$ within a galaxy, with parametric dependence on global galaxy properties $X$ (e.g. $M_*$, $M_\mathrm{gas}$, $R_\mathrm{d}$, $T$). The full functional form of the mass discrepancy, $\mathcal{D}(r; X)$, is a kinematic parametrization of the relation between dark matter and baryons, and a fortiori of the galaxy--halo connection.

While in principle $\mathcal{D}$ may have an arbitrary dependence on position, its utility is significantly enhanced by the fact that it is known to correlate strongly with acceleration $a(r)$ in all galaxies in which it has been measured in detail~\citep{Sanders_MDAcc, MG99, McGaugh_MDAcc, Tiret_Combes, Famaey_McGaugh, Janz}. This allows us to focus our attention on $\mathcal{D}(a(r); X)$, which is known as the ``mass discrepancy--acceleration relation'' or MDAR. The information content of the MDAR relative to one-point summaries may be approximately established by a simple counting argument: while the latter are limited to one data point per galaxy, the former contains as many data points as one can measure across a galaxy's entire rotation curve.

The aim of this paper is to construct a framework for using the MDAR to test galaxy formation models, and hence deduce the dynamically relevant correlations of the galaxy--halo connection. Although the MDAR has been known observationally for decades, few studies have sought to systematically extract the information contained within it. The tightness of the relation is used by some to argue against all $\Lambda$CDM-based models of galaxy formation (e.g.~\citealt{MG98, MG99, McGaugh_MDAcc, Kroupa_Falsification, TTP, Kroupa}), where stochasticity in the galaxy--halo connection may be expected to introduce significant scatter into the relation between $\mathcal{D}$ and $a$. In addition, it is argued that the presence of a ``characteristic acceleration'' $a \approx 10^{-10} \: \mathrm{m\:s}^{-2}$ at which $\mathcal{D}$ consistently becomes $\sim 1$ (indicating a dynamically insignificant quantity of dark matter) is not compatible with standard theory. A generic galaxy--halo connection would predict a spread in $\mathcal{D}$ at high $a$ and no clear transition in acceleration space between the dark matter and baryon-dominated regimes. Other authors, however, claim that the salient features of the MDAR arise naturally in $\Lambda$CDM models that have been tuned to match the Tully--Fisher~\citep{vdB} or $M_*-M_\mathrm{halo}$, $M_*-R_\mathrm{d}$, and $M_*-M_\mathrm{gas}$ relations~\citep{DC}, and consensus concerning the relation's significance does not seem at hand. No study to date has systematically investigated the dependence of $\mathcal{D}$ on global galaxy properties ($X$) at fixed $a$, or quantified the correlation of MDAR residuals with $r$.

Our specific task is twofold. First, we create a set of statistics to quantify four significant features of the MDAR: its shape, its scatter, the presence of a ``characteristic acceleration,'' and the correlation of its residuals with other variables ($M_*$, $M_\mathrm{gas}$, $R_\mathrm{d}$, $T$ and $r$). This is motivated in part by the prevalence of largely qualitative claims in the literature concerning the compatibility of the MDAR with various models, from which it may be difficult to determine the exact degree or nature of the agreement. Our statistics enable the conversion of verbal assertions into precise statistical comparisons, which we hope will sharpen discussion of the MDAR regardless of theoretical perspective.

The salience of these statistical features, however, is best appreciated in the context of specific model expectations. Our second task, therefore, is to develop a semi-empirical framework in $\Lambda$CDM to generate predictions for the MDAR. We adopt a fully bottom-up methodology, beginning with the simplest and best motivated correlations between galaxy and halo variables, and introducing more when required by the data. By comparing the predicted and observed MDARs, we will deduce the extent to which semi-empirical models are able to account for the significant aspects of the relation, and the concrete extensions to basic models that are required to match the relation's more detailed features. We intend in this way to lay the groundwork for a phenomenological determination of the galaxy--halo connection from information-rich relations such as the MDAR, as well as formulate precise tests for specific models.

The starting point of our framework is the technique of halo abundance matching (AM), which imposes a nearly monotonic relationship between galaxy stellar mass and halo mass or velocity at a particular epoch~\citep{Kravtsov,Conroy,Behroozi_2010,Guo,Moster}. From a phenomenological perspective, AM specifies the relation between stellar mass and halo mass and concentration required to fit clustering~\citep{Reddick, Lehmann} and dynamical~\citep{DW15, DW16} observations, but in its basic form neglects gas mass as well as galaxy size and type. We therefore augment the model by allowing correlations of these variables with $M_\mathrm{vir}$ and $c$ at fixed $M_*$. For a given set of correlations constituting the galaxy--halo connection, we generate a large number of mock data sets from our theoretical population with baryonic properties identical to the real data and halo properties specified by the model. We then calculate the MDAR statistics of these mock data sets, and compare with the observations.

The MDAR is considered by some an important piece of evidence in favour of Modified Newtonian Dynamics (MOND) as an alternative to $\Lambda$CDM for solving the missing mass problem, and is a central relation in MOND phenomenology (see~\citealt{Famaey_McGaugh} and references therein). Indeed, the founding papers of MOND were the first to predict that $\mathcal{D}$ would be more tightly correlated with acceleration than velocity or galactocentric radius~\citep{Milgrom1, Milgrom2, Milgrom3}, a hypothesis not verified empirically for many years. In MOND, the MDAR is a direct manifestation of the breakdown of Newtonian gravity or mechanics at low acceleration $a < a_0 \approx 1.2 \times 10^{-10} \: \mathrm{m\:s}^{-2}$, with the result that it is predicted to have negligible intrinsic scatter, residuals systematically uncorrelated with any other variable, and a clear acceleration scale $a=a_0$. While the MOND MDAR must go to 1 at $a \gg a_0$ (the Newtonian regime) and has a shape fixed by the theory at $a \ll a_0$ (the deep-MOND regime), the behaviour at intermediate $a$ is specified by an ad hoc interpolating function. Our statistical analysis will shed light on the compatibility of the observations with the MOND hypothesis, and our comparison with $\Lambda$CDM models will bring into sharper focus the relative evidence accorded to MOND and $\Lambda$CDM by the relation.

The structure of this paper is as follows. In Section~\ref{sec:data} we describe our observational MDAR sample and the $N$-body simulations on which we build our theoretical framework. In Section~\ref{sec:method} we lay out our procedure for constructing the galaxy--halo connection and document the MDAR statistics with which we evaluate our models. In Section~\ref{sec:results} we present our comparison of theory and data, first for a fiducial model best motivated by prior analyses, and then allowing variations in our model assumptions to maximise agreement with the MDAR. Section~\ref{sec:discussion} discusses literature studies in the context of our results, locates our parameter constraints and model requirements relative to previous findings, and elaborates on the broader implications for our understanding of galaxy astrophysics. Section~\ref{sec:conclusion} concludes.

\section{Observed and Simulated Data}
\label{sec:data}

\begin{figure}
  \centering
  \includegraphics[width=0.5\textwidth]{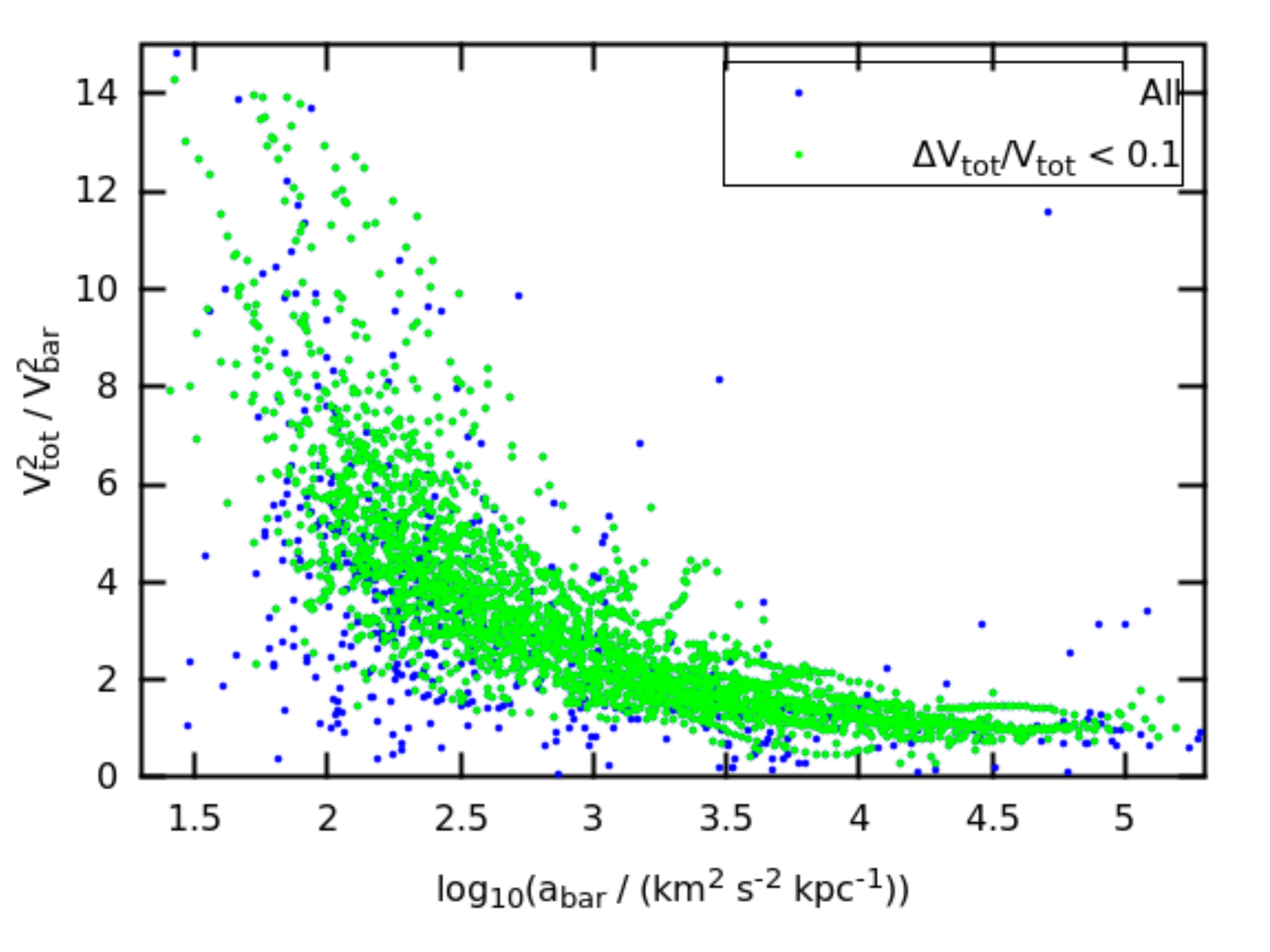}
  \caption{The mass discrepancy--acceleration relation of the high-quality subset of the \textsc{sparc} data. 153 galaxies are included with an average of 21 measurement points for each. The highest precision points, which are likely to be least affected by non-circular motions, are shown in green.}
  \label{fig:fig1}
\end{figure}

We take the observational MDAR from the \textsc{sparc} sample~\citep*{SPARC},\footnote{\url{http://astroweb.cwru.edu/SPARC/}} a recent compilation of $\sim 175$ high-quality resolved rotation curves augmented with \textit{Spitzer} $3.6 \: \upmu \mathrm{m}$ photometry. This sample spans a wide range in all dynamically relevant galaxy properties ($6.7 < \log(M_*/M_\odot) < 11.5$, $0.18 < R_\mathrm{d}/\mathrm{kpc} < 18.8$, $7.2 < \log(M_\mathrm{gas}/M_\odot) < 10.7$, and Hubble type S0 to Im/BCD), making it ideal for our analysis. Following the suggestion of~\citet{SPARC}, we remove galaxies with quality flag $Q=3$ (indicating major asymmetries, non-circular motions, and/or offsets between the stellar and H\textsc{i} distributions) and those with inclination $i<30^{\circ}$, which may be subject to large systematic uncertainties in their rotation curves. We take $M_\mathrm{gas}=1.33 \: M_\mathrm{HI}$ to account for cosmological helium, and adopt a stellar mass-to-light ratio of 0.5 for the disc and 0.7 for the bulge~\citep{SPARC}. The MDAR of the final sample of 153 galaxies is shown in Figure~\ref{fig:fig1}.

Our theoretical framework is based on dark matter haloes from the \textsc{darksky} simulation suite~\citep{DarkSky}. The \textsc{darksky} suite\footnote{\url{http://darksky.slac.stanford.edu/}} assumes a flat $\Lambda$CDM cosmology with $h=0.688$, $\Omega_\mathrm{m} = 0.295$, $n_\mathrm{s} = 0.968$, and $\sigma_8 = 0.834$. Here we use the \textsc{darksky-400} simulation (\verb|ds14_i_4096|), a $(400 \: \mathrm{Mpc~h^{-1}})^3$ box with $4096^3$ particles and mass resolution of $7.63 \times 10^7 \: h^{-1} M_\odot$, run with the \textsc{2hot} code~\citep{Warren13}. This simulation was previously used in the studies of~\citet{Lehmann},~\citet{DW16}, and~\citet{Jennings}. Haloes were identified using the \textsc{rockstar} halo finder~\citep*{Rockstar_1}, and a merger tree generated using the \textsc{consistent trees} code~\citep{Rockstar_2}.

\section{Method}
\label{sec:method}

We construct a series of empirical models of increasing complexity for the effective galaxy--halo connection of the \textsc{sparc} sample. We begin with the standard AM ansatz that the only galaxy property that correlates with halo mass and concentration is stellar mass. We then consider the possibility that the \textsc{sparc} galaxies inhabit a systematically biased subset of the total halo population, the potential for correlation between halo concentration and galaxy size at fixed stellar mass, and finally the impact of disc formation on the halo density profile. To test the models we use them to generate a large number of mock MDAR data sets with baryonic properties identical to the real data, but halo properties specified by the model. For each mock data set, we calculate a series of statistics based on important features of the MDAR, which we then compare to the values derived from the real data. We describe our theoretical framework in Section~\ref{sec:model} and our statistical analysis of the MDAR in Section~\ref{sec:comparison}. For quick reference, the free parameters of the framework are listed in Table~\ref{tab:params1}, and the MDAR statistics in Table~\ref{tab:params2}.

\begin{table*}
  \begin{center}
    \begin{tabular}{l|l}
      \hline
      $\alpha$				& Interpolates AM proxy between $M_\mathrm{vir}$ ($\alpha=0$) and $v_\mathrm{max}$ ($\alpha=1$)\\
      AM scatter			& Universal Gaussian scatter in stellar mass at fixed proxy\\
      $f$				& Fraction of high-concentration haloes removed at fixed $M_*$\\
      $m$				& Slope of galaxy size--halo concentration correlation at fixed $M_*$\\
      $\nu$				& Controls the degree of halo expansion ($-$) or contraction (+)\\
      \hline
    \end{tabular}
  \caption{Free parameters of the framework.}
  \label{tab:params1}
  \end{center}
\end{table*}

\subsection{Model}
\label{sec:model}

We begin by assigning to the $i^\mathrm{th}$ galaxy in a given mock data set the observed stellar mass $M_*$ and disc scale length $R_\mathrm{d}$ of the $i^\mathrm{th}$ \textsc{sparc} galaxy. We perturb the observed stellar mass by the measurement uncertainty for that galaxy to estimate its true stellar mass. We then use AM to determine the set of \textsc{darksky-400} haloes that may host a galaxy of that stellar mass. To allow the most general dependence of stellar mass on halo mass and concentration, we use the AM parametrization of~\citet{Lehmann} in which the proxy is specified by a continuous free parameter, $\alpha$:

\begin{equation}
\int^\infty_{M_*} \phi(M_*') \: \mathrm{d}M_*' = \int^\infty_{v_\alpha} n(v_\alpha') \: \mathrm{d}v_\alpha',
\end{equation}

\noindent where $\phi$ denotes the galaxy number density per unit $M_*$, $n$ denotes the halo number density per unit $v_\alpha$, and

\begin{equation}
v_\alpha \equiv v_\mathrm{vir} \left(\frac{v_\mathrm{max}}{v_\mathrm{vir}}\right)^\alpha
\end{equation}

\noindent evaluated at the epoch of peak halo mass. We allow also for a universal Gaussian scatter between $M_*$ and $v_\alpha$, which we term the AM scatter. By matching to luminosity and comparing AM mocks to clustering and satellite fraction measurements,~\citet{Lehmann} find $\alpha = 0.6 \pm 0.2$ and AM scatter = $0.16 \pm 0.04$, which we consider as priors on our analysis.

Our stellar mass function will be the single-S\'{e}rsic fit of~\citet{Bernardi_SMF}. We caution that due to differences in photometric reductions, mass-to-light ratios or choice of initial mass function (IMF), these stellar masses may not be fully consistent with those of the \textsc{sparc} data. A mismatch would imply that a given \textsc{sparc} galaxy would be given a slightly incorrect halo mass, although this effect is largely degenerate with $\alpha$. We discuss this issue further in Section~\ref{sec:problems}. We note also that the~\citet{Bernardi_SMF} stellar mass function (in addition to the halo mass function from the simulation) requires extrapolation to reach the faint end of the \textsc{sparc} sample, $M_* \approx 10^{6.7} M_\odot$. We consider the benefits of a large stellar mass range to outweigh potential systematic errors from this extrapolation, and hence do not cut the sample at the resolution limit. The results at the low-mass end may however be affected by improvements in the resolution of simulations and the depth of galaxy surveys.

We now select from the AM catalogue a halo mass and concentration for the $i^\mathrm{th}$ galaxy in the mock data set, by one of the following three methods:

\begin{enumerate}

\item{} In the simplest method, we randomly select a halo from the stellar mass bin of the galaxy in question. (Where binning in stellar mass is required, we take 50 uniform bins in the range $6.7 < \log(M_*/M_\odot) < 11.5$.) This assumes no restriction on the type of halo that a \textsc{sparc} galaxy may occupy.

\item{} Alternatively, we suppose that, due to their morphology or other properties, the \textsc{sparc} galaxies inhabit a subset of the \textsc{darksky-400} haloes that is systematically offset from the mean in dynamically relevant variables. This could obtain for example if late-type galaxies tend to inhabit less-massive or less-concentrated haloes than early-types, as suggested by~\citet{Wojtak},~\citet{DW15}, and~\citet{Mandelbaum} among others. To model this effect, we remove some fraction $f$ of the highest concentration haloes at each stellar mass, and randomly select a halo from the remainder:

\begin{equation}
\frac{N_\mathrm{i}(c>c_\mathrm{max,i})}{N_\mathrm{tot,i}} = f
\end{equation}

\noindent for stellar mass bin $i$. A larger fraction corresponds to a stronger selection effect for the \textsc{sparc} sample. Note that $M_\mathrm{vir}$ could equally well be used instead of concentration in this step, as the two are largely degenerate in their effect on $\mathcal{D}$. We choose concentration because the stronger correlation between $v_\alpha$ and $M_\mathrm{vir}$ in our AM model causes $c$ to have a larger spread at fixed $M_*$.

\item{} Finally, we consider the possibility that galaxy variables besides $M_*$ are relevant for determining the type of halo to which a galaxy belongs. The second parameter most commonly considered in the literature is colour (e.g.~\citealt{Puebla, Hearin_Watson, CAM}), but for our purposes the most important variable neglected by AM is galaxy size, as quantified by $R_\mathrm{d}$. This varies over two orders of magnitude in the \textsc{sparc} sample and has a significant direct effect on a galaxy's kinematics and hence its position on the MDAR. Not only is there no reason for a correlation between $R_\mathrm{d}$ and $c$ at fixed $M_*$ not to exist, but in fact it is motivated both by its presence in the angular momentum-based galaxy formation model of~\citet*{MMW} and by the suggestion of~\citet{DW16} that such a correlation may be required to match the small scatter of the Fundamental Plane. To capture the leading order effect of this relation on the MDAR, we introduce a toy model for correlating the radius residuals of the $M_*-R_\mathrm{d}$ relation with the concentration residuals of the $M_*-c$ relation. We assert that on average the concentration residual is a fixed multiple $m$ of the radius residual, with a fiducial 0.1 dex Gaussian scatter:

\begin{equation}
\Delta c \sim \mathcal{N}(m \cdot \Delta R_\mathrm{d}, 0.1),
\end{equation}

\noindent where

\begin{equation}
\Delta x \equiv \log(x) - \langle \log(x)|M_* \rangle.
\end{equation}

When $m<0$, larger galaxies tend to live in less-concentrated haloes (in accordance with~\citealt{MMW}), and vice versa for $m>0$. This model will allow us to test for such a correlation in the MDAR data and absorb any discrepancies that may be related to the galaxy--halo connection in this way. When using this method, we first determine $c$ from $M_*$ and $R_\mathrm{d}$, and then randomly select $M_\mathrm{vir}$ from the subset of the AM-matched catalogue at this $M_*$ and $c$.

\end{enumerate}

We have now associated a halo with each \textsc{sparc} galaxy in a given mock data set; it remains to lay out the baryonic and dark matter mass profiles and determine $a$ and $\mathcal{D}$ along the rotation curve. We consider at this stage the impact of disc formation on the halo density distribution. While pristine dark matter haloes have a density profile well-characterised by the NFW form, galaxy formation draws dark matter inwards by adiabatic contraction~\citep{Blumenthal, Gnedin_2004, Gnedin_2011}, and may subsequently move it outwards again through stellar feedback (e.g.~\citealt{Mash, Governato, Pontzen}). Following~\citet{D07} and~\citet{DW15,DW16}, we parametrize these effects by means of a single free parameter $\nu$, which interpolates between the adiabatic contraction model of~\citet{Gnedin_2011} ($\nu=1$) and an expansion of the same magnitude ($\nu=-1$).\footnote{In some feedback models, dark matter is moved outwards only in the inner regions of galaxies, and continues to move in by adiabatic contraction further out. Since $\nu<0$ implies halo expansion at all radii, it may underestimate the amount of dark matter at intermediate to large radii relative to such models.} We take our fiducial value of $\nu$ to be $0$, corresponding to pure-NFW haloes. Although this parametrization must capture the leading order effect of halo expansion or contraction, we caution that in the absence of consensus concerning the effect of feedback on galaxy formation its precise mass and radius dependence remains somewhat arbitrary. We will discuss this issue further in Section~\ref{sec:problems}. $\nu$ is the fifth and final free parameter of our framework.

This prescription for halo contraction or expansion requires as input the full baryonic mass distribution of a model galaxy. To estimate this, we assume that each galaxy is composed of a small bulge, a thin exponential stellar disc, and a thin exponential gas disc. We take the bulge mass fraction, gas mass and stellar disc scale length directly from the \textsc{sparc} data on a galaxy-by-galaxy basis (the first by private communication, F. Lelli), and calculate the gas disc scale length $R_\mathrm{d,gas}$ from the measured radius ($R_\mathrm{HI}$) at which the H\textsc{i} surface density drops below $1 \: M_\odot \: \mathrm{pc}^{-2}$.\footnote{For some galaxies in which the gas profile is far from exponential it is not possible to derive a self-consistent value for $R_\mathrm{d,gas}$. In these cases we take the value closest to providing a solution. For galaxies in which $R_\mathrm{HI}$ was not measured, we use $R_\mathrm{d,gas} = 2R_\mathrm{d}$~\citep{SPARC}.} For a given value of $\nu$, and for a given abundance-matched galaxy, this enables us to determine the full dark matter mass profile $M_\mathrm{DM}(<r)$.

We are now in position to derive the MDAR of each mock data set. To do this, we calculate for the $i^\mathrm{th}$ mock galaxy the enclosed halo mass within each of the radii actually sampled for the $i^\mathrm{th}$ \textsc{sparc} galaxy, and hence derive $V^2_\mathrm{DM}$. We take the baryonic contribution $V_\mathrm{bar}^2$ to be exactly as calculated for the real data, and hence determine the total velocity $V^2 = V^2_\mathrm{bar} + V^2_\mathrm{DM}$. Since this is the ``true'' model velocity while the data use the observed velocity, we scatter it by the median quoted measurement uncertainty (7 per cent) before calculating the mass discrepancy $\mathcal{D}(r) = (V^2_\mathrm{bar}(r) + V^2_\mathrm{DM}(r))/V^2_\mathrm{bar}(r)$. We cast the MDAR in terms of the baryonic acceleration $a_\mathrm{bar} = V_\mathrm{bar}^2/r$ rather than the total acceleration $a_\mathrm{tot} = V_\mathrm{tot}^2/r$ so that the values of the independent variables in the mock and real data sets are identical. For convenience, we use the definition $a \equiv \log(a_\mathrm{bar}/(\mathrm{km}^2 \: \mathrm{s}^{-2} \: \mathrm{kpc}^{-1}))$ hereafter.

Note that we do \emph{not} use bulge--disc decomposition of the baryon mass profile in the calculation of $V_\mathrm{bar}^2$, which we instead take directly from the data: this decomposition is approximate and required \emph{only} for implementing our halo expansion/contraction scheme. Using identical $V_\mathrm{bar}$ values for the mock and real data sets ensures that any differences in their MDARs are due entirely to the dark matter distribution. Deviation of galaxies' baryonic mass profiles from the bulge--disc approximation does introduce a small error into $M_\mathrm{DM}(<r)$ for given $\nu$, but since halo expansion is a second-order effect in the determination of $\mathcal{D}$ -- and variations in the baryonic mass distribution are in any case largely degenerate with $\nu$ -- this is permissible.

By compiling the $a-\mathcal{D}$ curves for each of the 153 galaxies in a given mock data set, we produce a theoretical MDAR fully commensurable with that observed. We repeat this procedure 2000 times to generate an ensemble of mock MDAR data sets statistically representative of the underlying model, which we then subject to the following testing.

\subsection{Statistical analysis and comparison of theory and observation}
\label{sec:comparison}

As discussed in Section~\ref{sec:intro}, we wish to understand four aspects of the MDAR and their relation to theoretical expectations: shape, scatter, the presence of a ``characteristic acceleration'' beyond which mass discrepancies consistently go to $\sim 1$, and the correlation of residuals with other galaxy properties. We devise a series of statistics to quantify these features, and compare the observed values to the distributions obtained from Monte Carlo realisations of the models. In determining appropriate statistics, it is important to consider the level of detail at which one wishes to probe the relation. One cannot expect the model to match statistics that reflect finer-grained aspects of the relation than one can reasonably hope it to predict, but on the other hand statistics not fine enough may imply agreement where in detail there is none. In addition, we consider statistics explicitly tied to physically meaningful features of the MDAR to be more useful than those giving simply an overall measure of agreement, since they reveal the precise ways in which a model performs well or poorly.

We propose the following (summarised in Table~\ref{tab:params2}). (i)--(v) relate to the shape and scatter of the MDAR, (vi)--(xi) to the presence of a ``characteristic acceleration scale,'' and (xii) to the correlation of MDAR residuals with other galaxy properties. Where binning in acceleration is required, we use 12 bins uniformly spaced in the range $1.5 < a < 5$, so that the first and last bins occupy the ranges $1.5 < a < 1.8$ and $4.7 < a < 5$, respectively. We have checked that the details of the binning do not affect our conclusions.

\begin{table*}
  \begin{center}
    \begin{tabular}{l|l}
      \hline
      $\langle \mathcal{D} \rangle_{1.5-1.8}$			& Average $\mathcal{D}$ in range $1.5 < a < 1.8$\\
      $\langle \mathcal{D} \rangle_{4.7-5}$			& Average $\mathcal{D}$ in range $4.7 < a < 5$\\
      $\sigma(\mathcal{D})_{1.5-1.8}$				& Standard deviation in $\mathcal{D}$ over range $1.5 < a < 1.8$\\
      $\sigma(\mathcal{D})_{4.7-5}$				& Standard deviation in $\mathcal{D}$ over range $4.7 < a < 5$\\
      $\sigma_{\mathrm{tot}}$					& Weighted average standard deviation in $\mathcal{D}$ over range $1.5 < a < 5$\\
      $\langle \textgoth{a} \rangle$				& Average value of $a$ at which $\mathcal{D}$ drops below 3\\
      $\sigma(\textgoth{a})$					& Standard deviation in value of $a$ at which $\mathcal{D}$ drops below 3\\
      $\mathcal{F}_+$						& Number of galaxies with $\mathcal{D} > 3$ across their rotation curve\\
      $\mathcal{F}_-$						& Number of galaxies with $\mathcal{D} < 3$ across their rotation curve\\
      $\langle a \rangle_{2-2.5}$       			& Average $a$ for points with $2 < \mathcal{D} < 2.5$\\
      $\sigma(a)_{2-2.5}$					& Standard deviation in $a$ for points with $2 < \mathcal{D} < 2.5$\\
      $\langle \Delta(\mathcal{D})_{M_*} \rangle$		& Weighted average difference in $\mathcal{D}$ between high and low $M_*$ galaxies at fixed $a$\\
      $\langle \Delta(\mathcal{D})_{\Delta R_\mathrm{d}} \rangle$		& Weighted average difference in $\mathcal{D}$ between large and small galaxies at fixed $M_*$ and $a$\\
      $\langle \Delta(\mathcal{D})_{\Delta M_\mathrm{gas}} \rangle$	& Weighted average difference in $\mathcal{D}$ between high and low $M_\mathrm{gas}$ galaxies at fixed $M_*$ and $a$\\
      $\langle \Delta(\mathcal{D})_T \rangle$	& Weighted average difference in $\mathcal{D}$ between late and early type galaxies at fixed $a$\\
      $\langle \Delta(\mathcal{D})_r \rangle$			& Weighted average difference in $\mathcal{D}$ between high and low sampling radii at fixed $a$\\
      \hline
    \end{tabular}
  \caption{MDAR diagnostic statistics. Rows 1--5 measure the shape and scatter of the MDAR, rows 6--11 quantify the relation's ``acceleration scale,'' and rows 12--16 assess the dependence of $\mathcal{D}$ on various dynamically relevant galaxy properties at fixed $a$. For the full definitions, see Eqs. 6--16.}
  \label{tab:params2}
  \end{center}
\end{table*}

\begin{enumerate}

\item{} \begin{equation} \langle \mathcal{D} \rangle_{1.5-1.8} \equiv \frac{\sum_{a=1.5}^{1.8} \mathcal{D}}{N_a(1.5, 1.8)} \end{equation}

\item{} \begin{equation} \langle \mathcal{D} \rangle_{4.7-5} \equiv \frac{\sum_{a=4.7}^{5} \mathcal{D}}{N_a(4.7,5)}, \end{equation}

\noindent where $N_Y(x,y)$ is the number of data points in the range $x < Y < y$, for $Y \in \{a, \mathcal{D}\}$. Statistic (i) describes the average relative amount of dark and baryonic matter in low acceleration regions, and (ii) is the same quantity in high-acceleration regions.

\item{} \begin{equation} \sigma(\mathcal{D})_{1.5-1.8} \equiv \left(\frac{\sum_{a=1.5}^{1.8} \mathcal{D}^2}{N_a(1.5, 1.8)} - \langle \mathcal{D} \rangle_{1.5-1.8}^2\right)^\frac{1}{2} \end{equation}

\item{} \begin{equation} \sigma(\mathcal{D})_{4.7-5} \equiv \left(\frac{\sum_{a=4.7}^{5} \mathcal{D}^2}{N_a(4.7,5)} - \langle \mathcal{D} \rangle_{4.7-5}^2\right)^\frac{1}{2} \end{equation}

\item{} \begin{equation} \sigma_{\mathrm{tot}} \equiv \left(\frac{\sum_{i=1}^{12} \sigma(\mathcal{D})_i^2 \cdot N_a(a_i, a_{i+1})}{N_a(1.5, 5)}\right)^\frac{1}{2}, \end{equation}

\noindent where $i$ is the bin index and $a_i$ is the lower limit of bin $i$. This is the average standard deviation in $\mathcal{D}$ over all the bins weighted by the number of points in each bin, and measures the overall scatter of the MDAR.

\item{} \begin{equation} \langle \textgoth{a} \rangle \equiv \frac{\sum_{j=1}^{N_\mathrm{cross}} \textgoth{a}_j}{N_\mathrm{cross}}, \end{equation}

\noindent where $\textgoth{a}_j$ is the $a$ value at which galaxy $j$ crosses $\mathcal{D}=3$, and $N_\mathrm{cross} = 153 - \mathcal{F}_+ - \mathcal{F}_-$ is the number of galaxies within this data set for which this happens (see below). We choose a threshold value of 3 because this minimises $\mathcal{F}_+ + \mathcal{F}_-$ in the data, and hence maximises the number of galaxies for which $\textgoth{a}$ is defined. Moderate perturbations to the threshold value ($\pm 1$) do not significantly affect the results. To calculate $\textgoth{a}$, we linearly interpolate between the two measured accelerations on either side of $\mathcal{D}=3$.

$\textgoth{a}_j$ is a proxy for the acceleration at which dark matter becomes dynamically negligible in galaxy $j$, and hence $\langle \textgoth{a} \rangle$ indicates the ``characteristic acceleration scale'' of this data set. Note that in a given data set not all galaxies have some measured points at $\mathcal{D} > 3$ and some at $\mathcal{D} < 3$; for galaxies that do not $\textgoth{a}$ is undefined. The number of such galaxies is measured by $\mathcal{F}_+$ and $\mathcal{F}_-$ (statistics (viii) and (ix) below).

\item{} \begin{equation} \sigma(\textgoth{a}) \equiv \left(\frac{\sum_{j=1}^{N_\mathrm{cross}} \textgoth{a}_j^2}{N_\mathrm{cross}} - \langle \textgoth{a} \rangle^2\right)^\frac{1}{2}. \end{equation}

\noindent The spread in $\textgoth{a}$ values among galaxies measures the similarity of the ``acceleration scales'' of different galaxies within a data set. The smaller $\sigma(\textgoth{a})$, the more unique this scale and hence the more it may be said to be ``characteristic.''

\item{} $\mathcal{F}_+$, the number of galaxies with $\mathcal{D} > 3$ across their rotation curve (typically low surface brightness galaxies). These do not possess a value for $\textgoth{a}$ and are therefore excluded in the calculation of $\langle \textgoth{a} \rangle$ and $\sigma(\textgoth{a})$.

\item{} $\mathcal{F}_-$, the number of galaxies with $\mathcal{D} < 3$ across their rotation curve (typically high surface brightness galaxies without measurements at large $r$). As above.\footnote{Note that a fair comparison of $\mathcal{F}_+$ and $\mathcal{F}_-$ between mock and real data sets requires that the rotation curves be sampled at identical radii for each. Analysis at this level of detail could not therefore be performed using the methods of~\citet{vdB} or~\citet{DC}, for example.}

\item{} \begin{equation} \langle a \rangle_{2-2.5} \equiv \frac{\sum_{\mathcal{D}=2}^{2.5} a}{N_\mathcal{D}(2, 2.5)}. \end{equation}

\noindent This is an alternative measure of characteristic acceleration. Given the typical overall measurement error in $\mathcal{D}$ ($\sim 35$ per cent; private communication, F. Lelli), it is at $2 < \mathcal{D} < 2.5$ that galaxies are likely to begin transitioning into the fully baryon-dominated regime.

\item{} \begin{equation} \sigma(a)_{2-2.5} \equiv \left(\frac{\sum_{\mathcal{D}=2}^{2.5} a^2}{N_\mathcal{D}(2, 2.5)} - \langle a \rangle_{2-2.5}^2\right)^\frac{1}{2}, \end{equation}

\noindent the associated spread. Statistics (x) and (xi) have the advantage over (vi) and (vii) that no galaxies need to be excluded, but the disadvantage that the information they contain is not specific to individual galaxies.

\item{} \begin{align}
\langle &\Delta(\mathcal{D})_X \rangle \equiv \\ \nonumber
&\frac{\sum_{i=1}^{12} (\langle \mathcal{D} \rangle_{i; X_2} - \langle \mathcal{D} \rangle_{i; X_1}) \cdot \sqrt{N_{a;X_1}(a_i, a_{i+1}) N_{a;X_2}(a_i, a_{i+1})}}{\sum_{i=1}^{12} \sqrt{N_{a;X_1}(a_i, a_{i+1}) N_{a;X_2}(a_i, a_{i+1})}},
\end{align}

\noindent where

\begin{equation} 
X \in \{M_*, \Delta R_\mathrm{d}, \Delta M_\mathrm{gas}, T, r\},
\end{equation}

\noindent $\Delta x$ is defined in Eq. 5, $X_1$ labels points at low $X$, and $X_2$ at high $X$. For $X \in \{M_*, \Delta R_\mathrm{d}, \Delta M_\mathrm{gas}\}$, $X_1$ corresponds to galaxies with $X$ in the bottom third of the \textsc{sparc} sample, and $X_2$ to galaxies with $X$ in the upper third. For $X=r$, $X_1$ corresponds to the third of measurement points taken at lowest galactocentric radius (separately for each galaxy), and $X_2$ to the corresponding upper third. For $X=T$, $X_1$ denotes galaxies with Hubble type S0--Sb (27 galaxies), and $X_2$ those with type Sd--Sdm (26 galaxies). We weight by $\sqrt{N_{a;X_1}(a_i, a_{i+1}) N_{a;X_2}(a_i, a_{i+1})}$ to prioritise the bins with the lowest shot noise, which contain many points from both subsamples.

These statistics measure the average vertical offset between the MDARs at high and low $X$. An offset significantly different from 0 would imply that the MDAR is dependent on $X$, and a fortiori that $\mathcal{D}$ is not a function of $a$ alone. The values of $X$ we consider here are the dynamically relevant galaxy properties that may be expected to influence the MDAR in a generic theory of galaxy formation. In particular, by comparing the predicted and observed $\langle \Delta(\mathcal{D})_X \rangle$ we can hope to get a handle on the following: the dynamical validity of the AM stellar mass--halo mass relation ($X=M_*$), the correlation of galaxy size with halo properties at fixed stellar mass ($X=\Delta R_\mathrm{d}$), the validity of the assumption that the halo proxy in AM is uncorrelated with cold gas mass at fixed stellar mass ($X=\Delta M_\mathrm{gas}$), the relation between galaxy morphology and halo properties ($X=T$), and finally the radial dependence of the halo density distribution ($X=r$).

\end{enumerate}

Associated with each statistic for a given model is a two-tailed $p$-value equal to the fraction of mock data sets with a more extreme value for that statistic than the data. We will say that a model fails to account for statistic $j$ (and hence the associated feature of the MDAR) if a value of $j$ at least as extreme as the real data's has a $p<0.05$ probability of being randomly drawn from the mock data distribution. This method is an example of Approximate Bayesian Computation (e.g.~\citealt{ABC}).

\section{Results}
\label{sec:results}

\subsection{Fiducial model}
\label{sec:fiducial}

\begin{figure}
  \centering
  \includegraphics[width=0.5\textwidth]{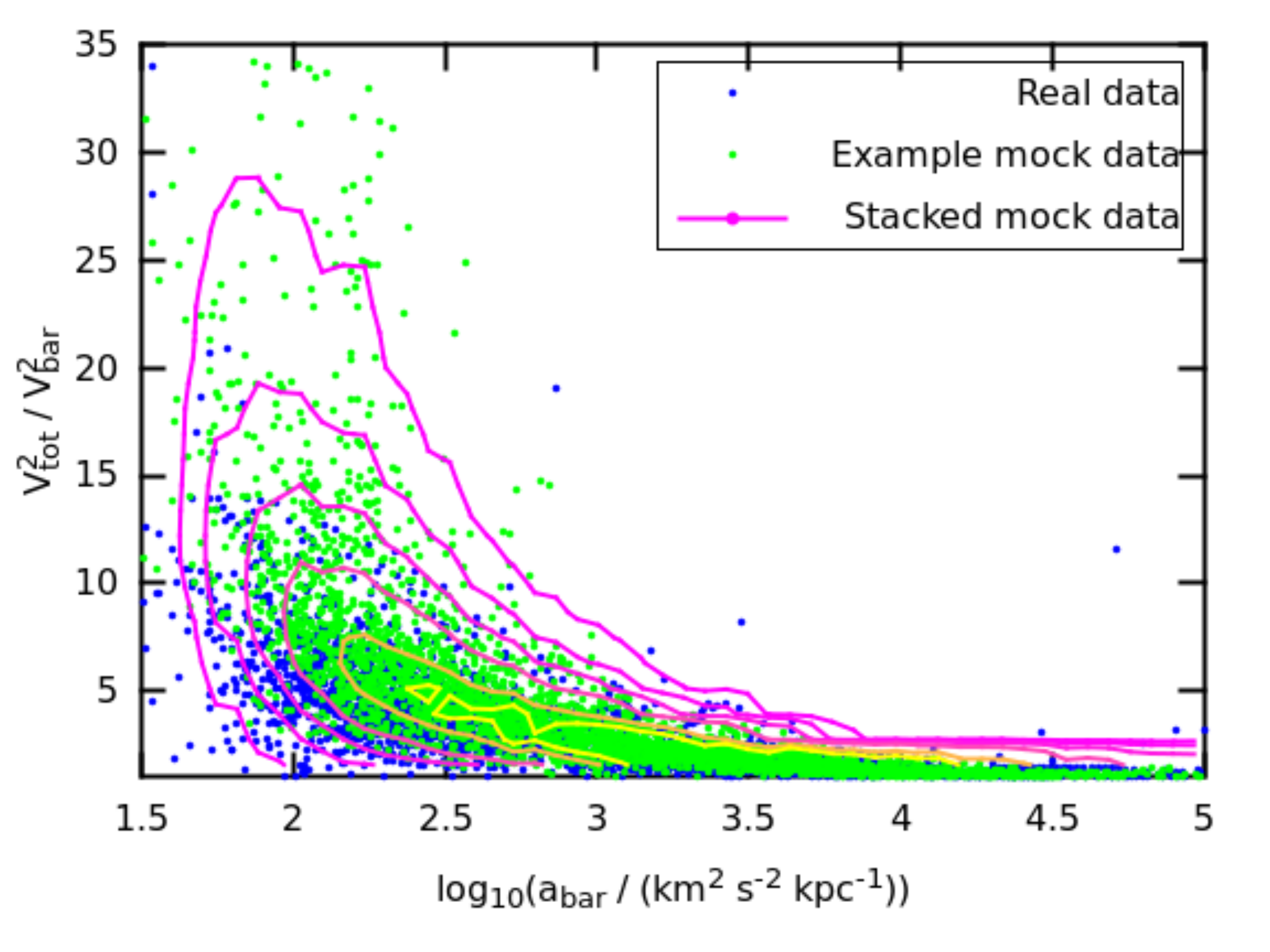}
  \caption{Comparison of the observed MDAR (blue points) to that predicted by the fiducial model (green points and contour). The contours enclose 40, 60, 80, 90, 95, and 98 per cent of the data points from 2000 Monte Carlo realisations of the model, while the green points show a randomly chosen realisation.}
  \label{fig:fig2}
\end{figure}

\begin{figure*}
  \subfigure[]
  {
    \includegraphics[width=0.48\textwidth]{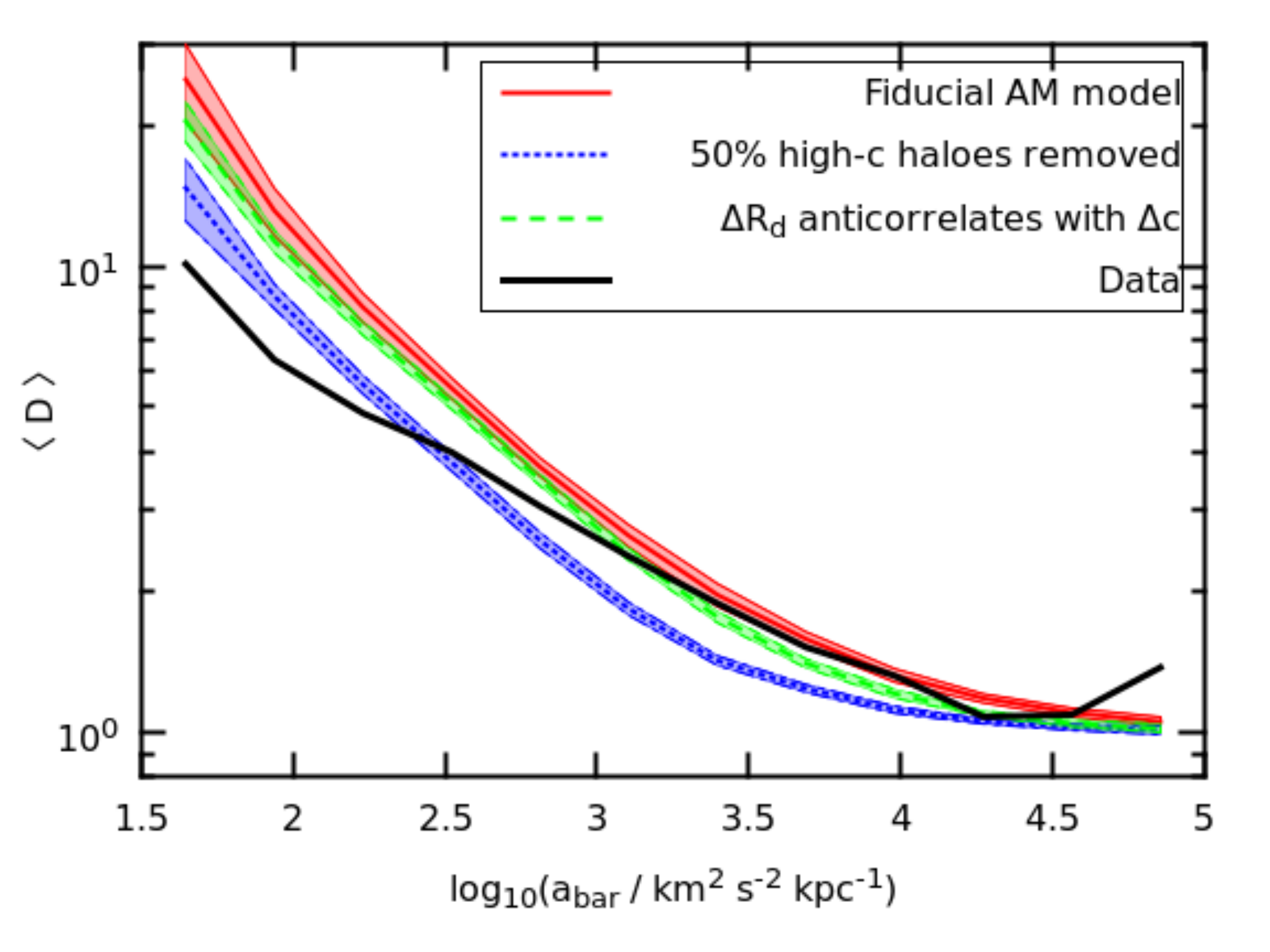}
    \label{fig:fig3a}
  }
  \subfigure[]
  {
    \includegraphics[width=0.48\textwidth]{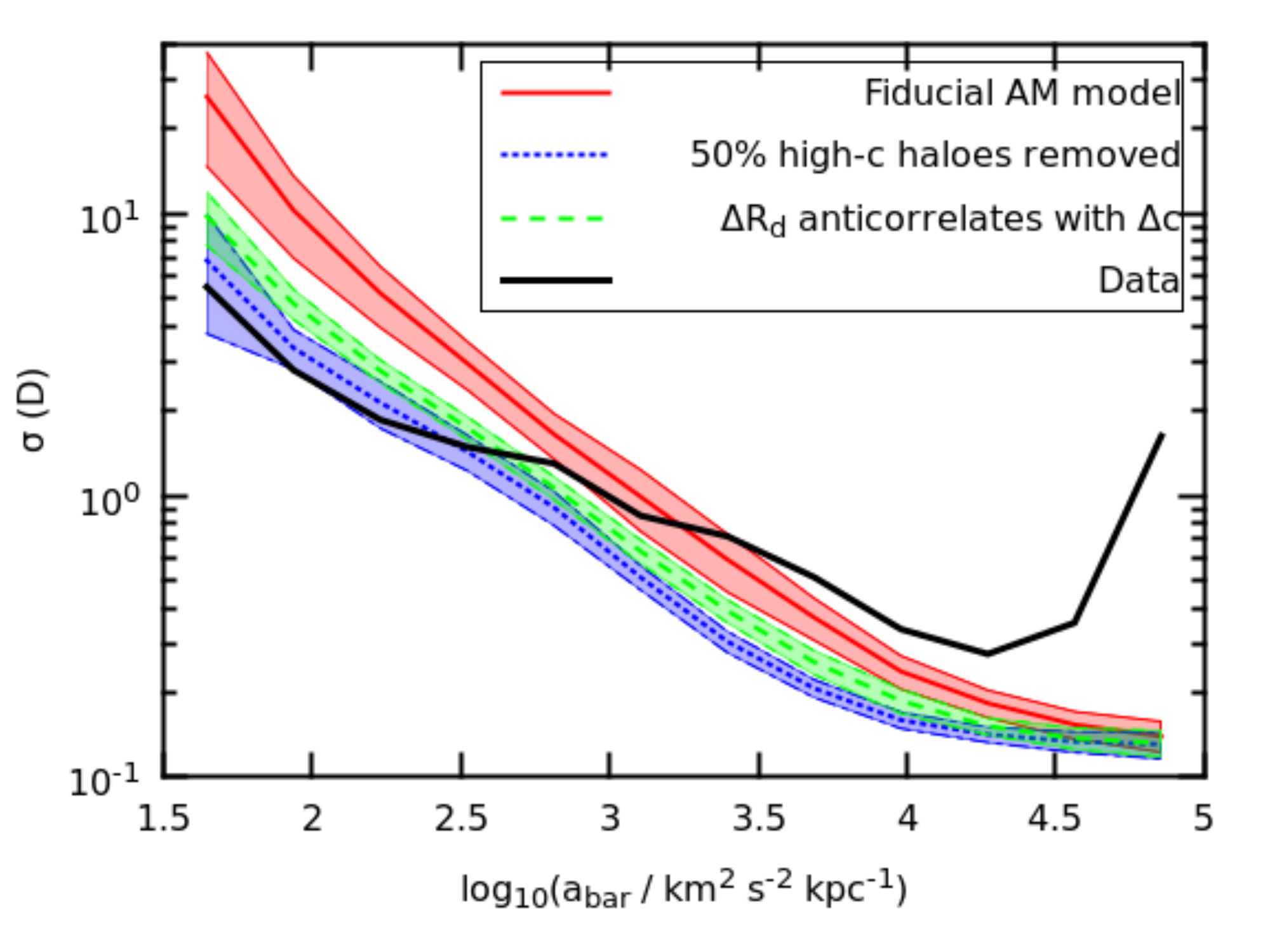}
    \label{fig:fig3b}
  }
  \caption{The mean (lines) and $1 \sigma$ scatter (shaded bands) across 2000 mock data sets of the mean mass discrepancy ($\langle \mathcal{D} \rangle$; Fig.~\ref{fig:fig3a}) and its standard deviation ($\sigma(\mathcal{D})$; Fig.~\ref{fig:fig3b}) in 12 bins of acceleration, compared to the real data values. The red, blue, and green lines correspond to rows two, seven, and nine in Tables~\ref{tab:table1}--\ref{tab:table3}, respectively. Only a strong selection on halo concentration can reduce the predicted $\langle \mathcal{D} \rangle$ and $\sigma(\mathcal{D})$ at low $a$ to acceptable levels.}
  \label{fig:fig3}
\end{figure*}

\begin{figure}
  \centering
  \includegraphics[width=0.5\textwidth]{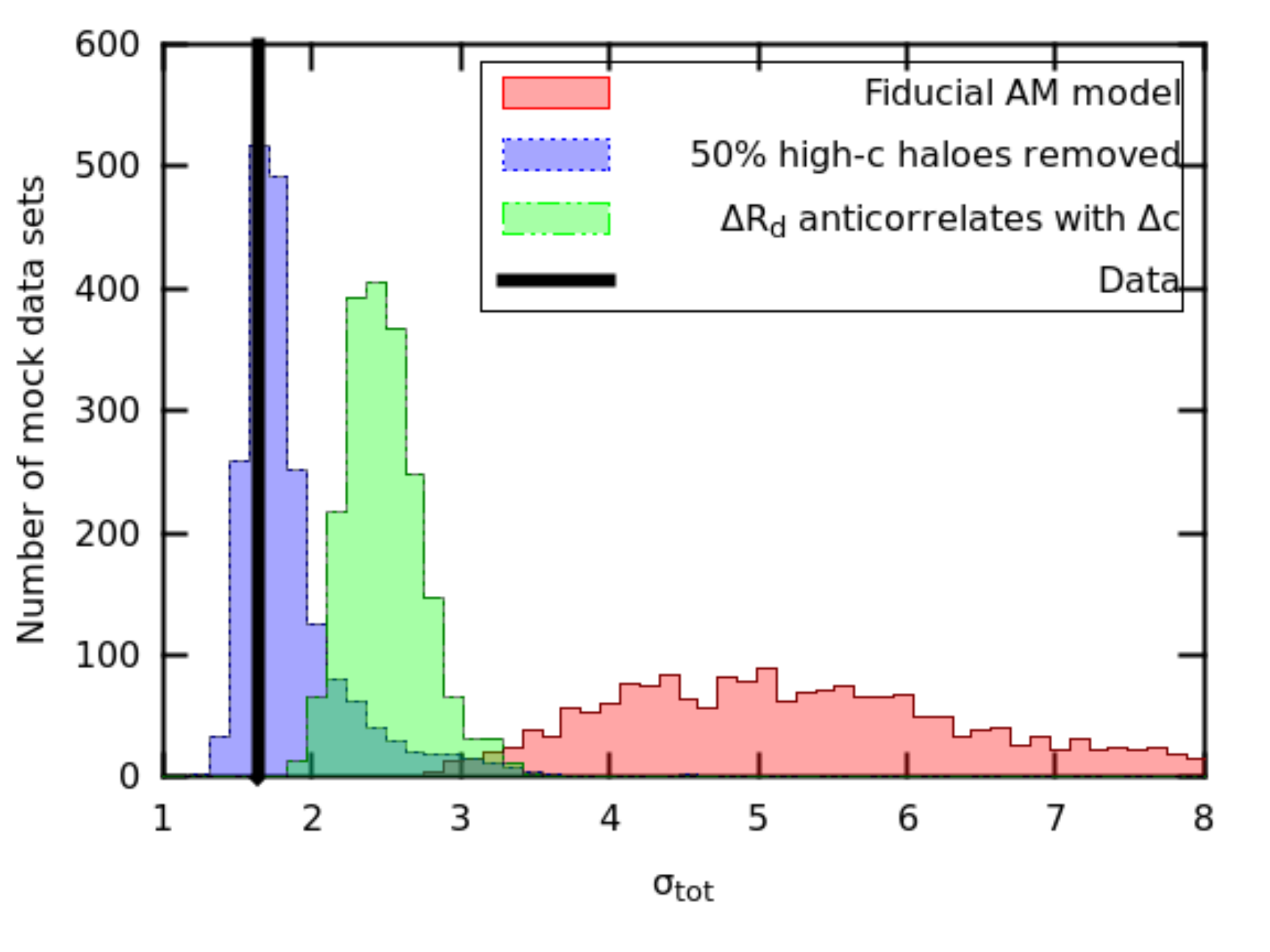}
  \caption{Distributions of the weighted average standard deviation ($\sigma_\mathrm{tot}$) in the MDAR for various model assumptions, compared to the value in the data. Only a strong selection on halo concentration can reduce the predicted $\sigma_\mathrm{tot}$ to acceptable levels.}
  \label{fig:fig4}
\end{figure}

\begin{figure*}
  \subfigure[Fiducial]
  {
    \includegraphics[width=0.48\textwidth]{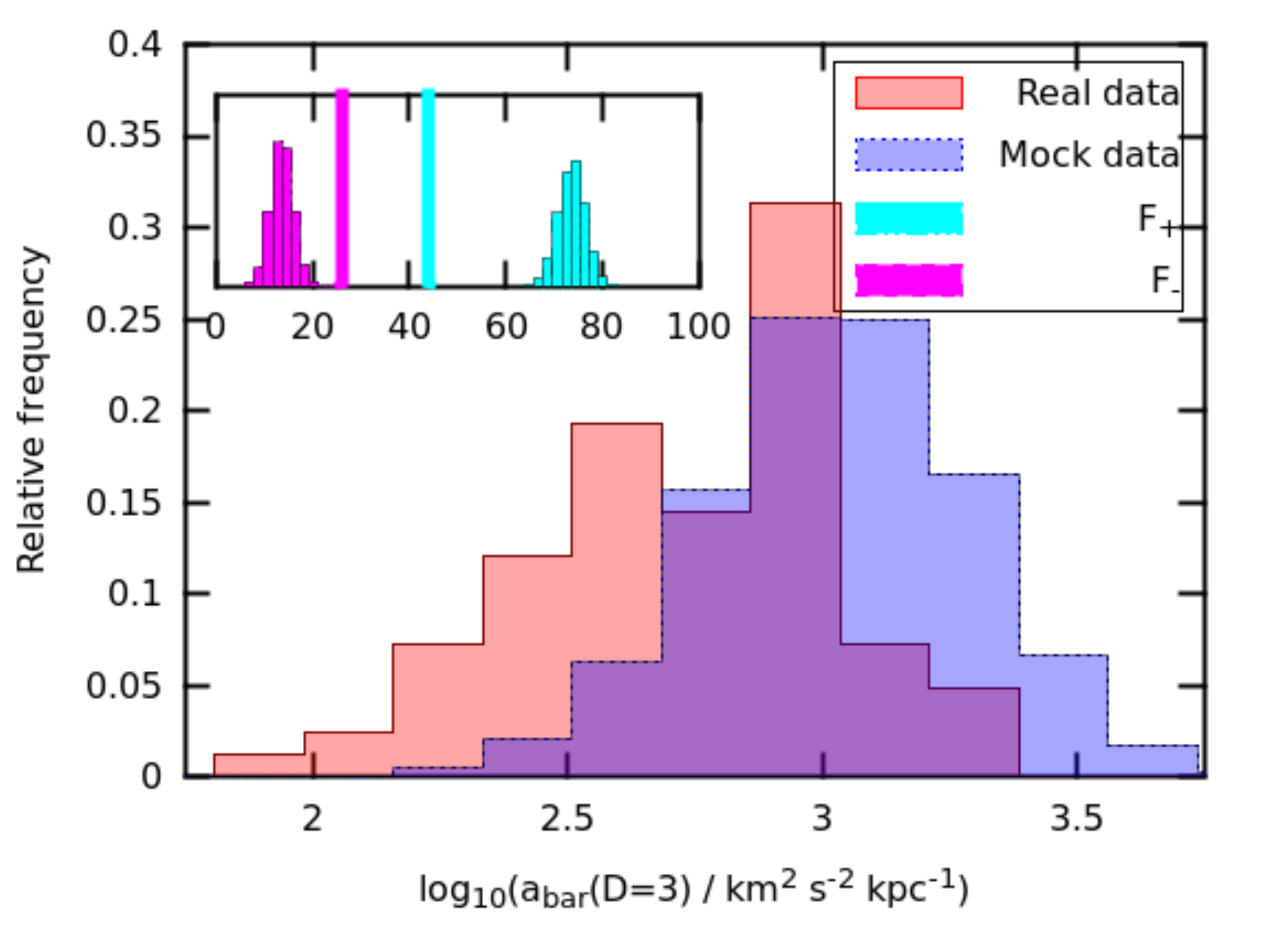}
    \label{fig:fig5a}
  }
  \subfigure[$\nu=-1$]
  {
    \includegraphics[width=0.48\textwidth]{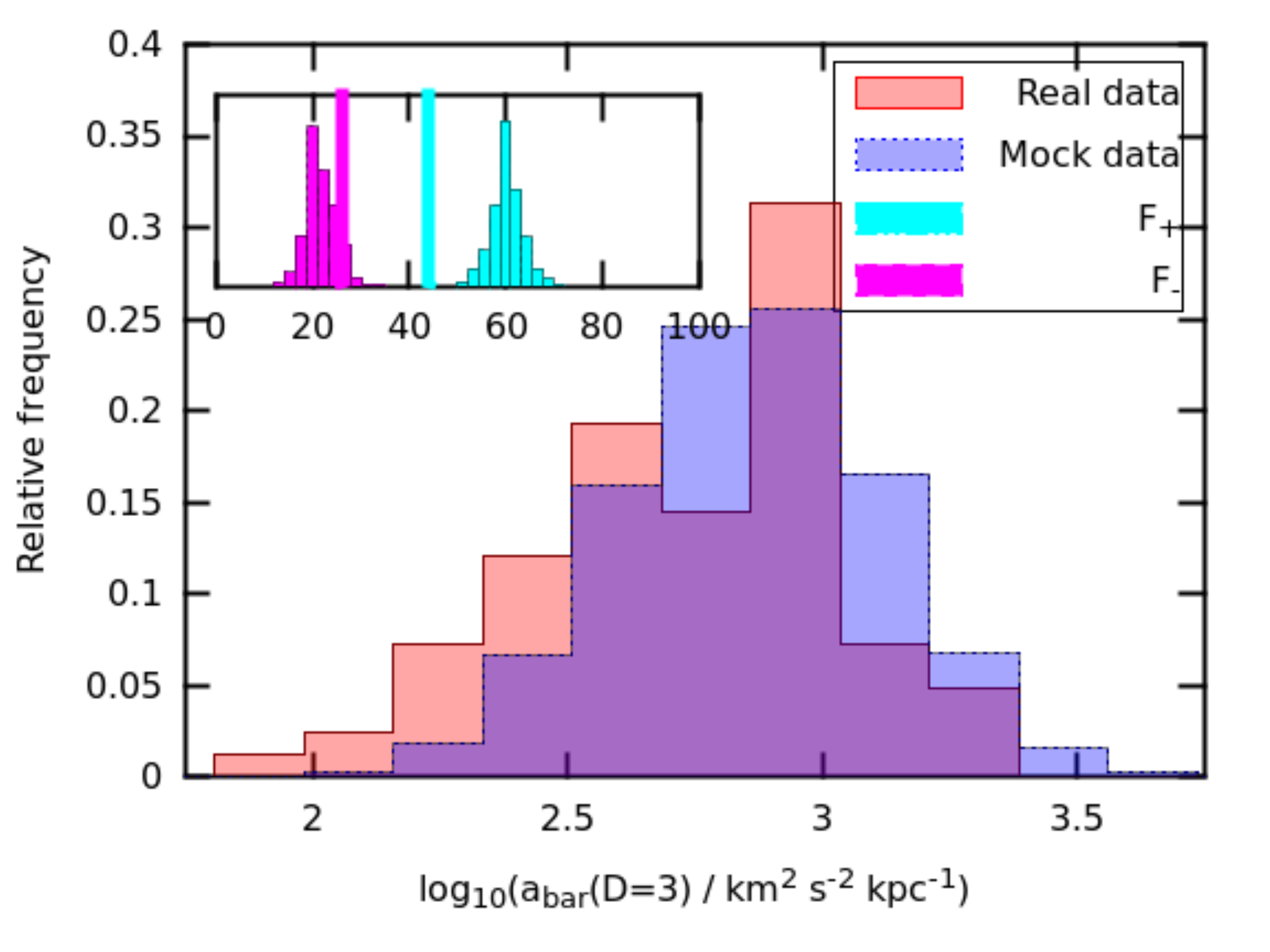}
    \label{fig:fig5b}
  }
  \subfigure[$f=0.5$]
  {
    \includegraphics[width=0.48\textwidth]{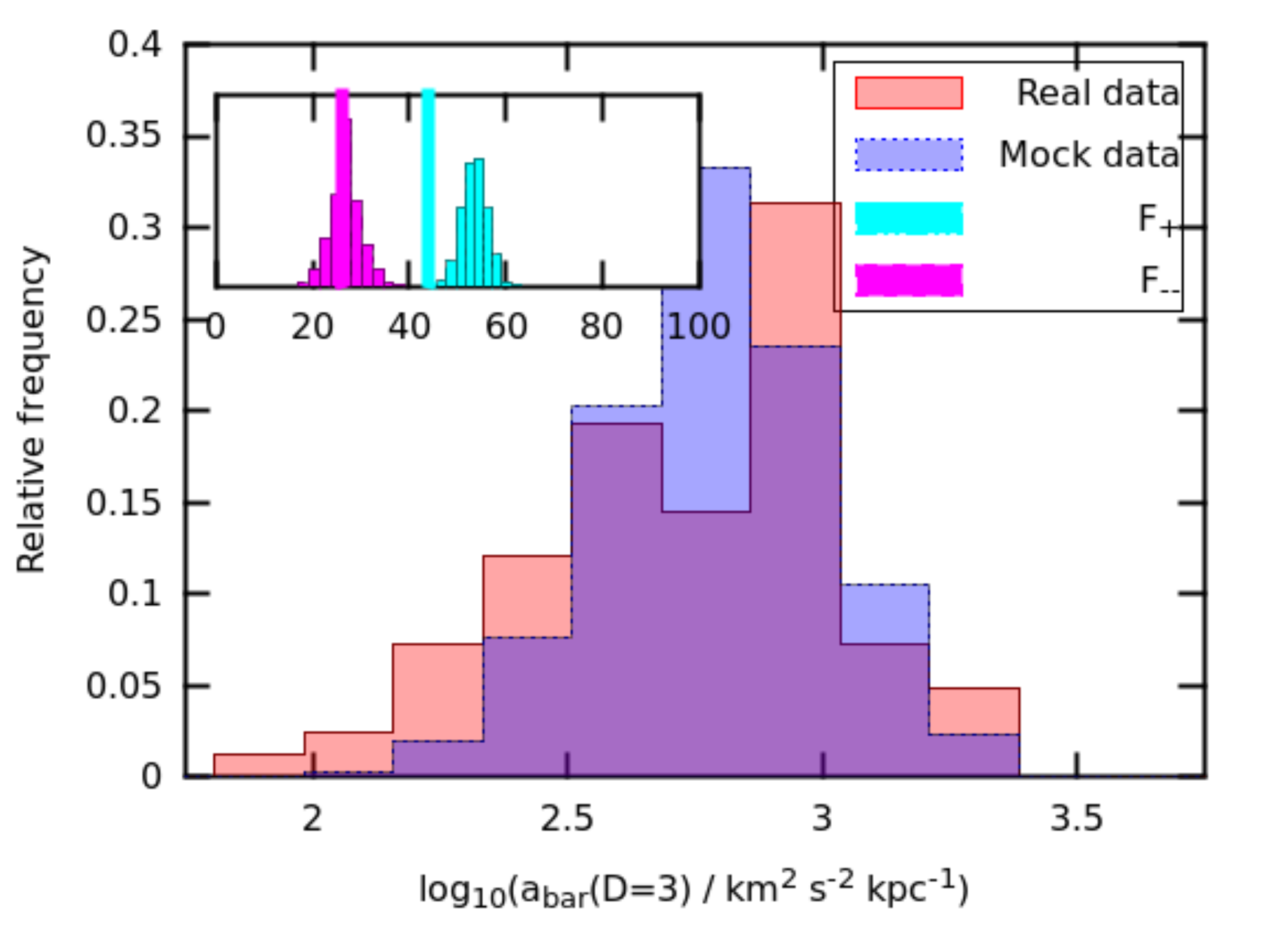}
    \label{fig:fig5c}
  }
  \subfigure[$m=-0.4$]
  {
    \includegraphics[width=0.48\textwidth]{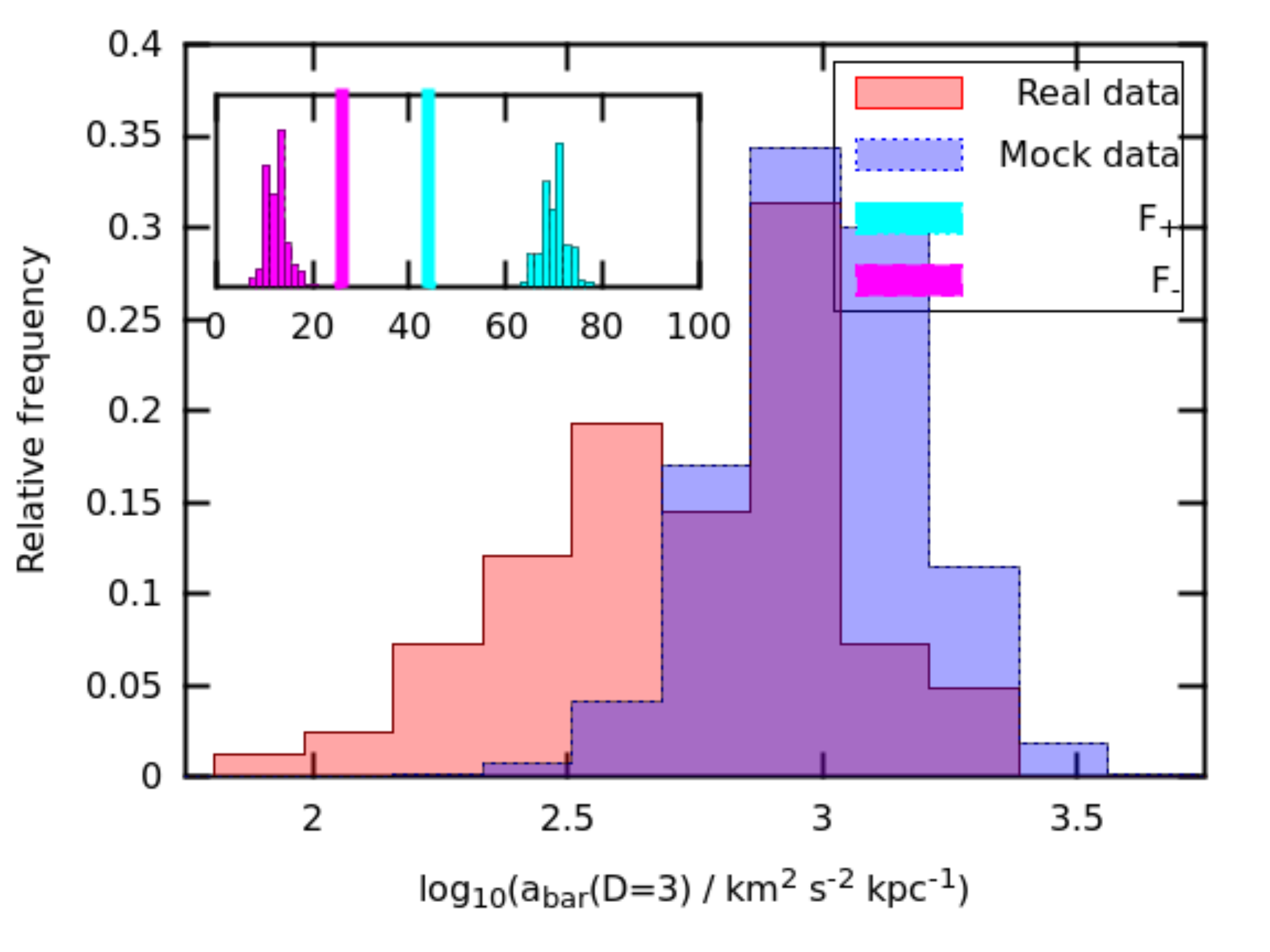}
    \label{fig:fig5d}
  }
  \caption{The observed (red) and predicted (blue) values of the acceleration (denoted by $\textgoth{a}$ in the text) at which the mass discrepancy crosses $\mathcal{D}=3$. This is a proxy for the ``characteristic acceleration'' at which dark matter becomes unimportant. The blue histograms combine the results of all 2000 Monte Carlo realisations of the model. The inset shows the number of galaxies in each realisation with all measured points above $\mathcal{D}=3$ (cyan; $\mathcal{F}_+$), and below $\mathcal{D}=3$ (magenta; $\mathcal{F}_-$), compared to the values in the real data (vertical lines). These galaxies are excluded from the main histograms. The four panels correspond to the models in rows two, four, seven, and nine of Table~\ref{tab:table2}, respectively.}
  \label{fig:fig5}
\end{figure*}

\begin{figure*}
  \subfigure[$\langle \textgoth{a} \rangle$]
  {
    \includegraphics[width=0.48\textwidth]{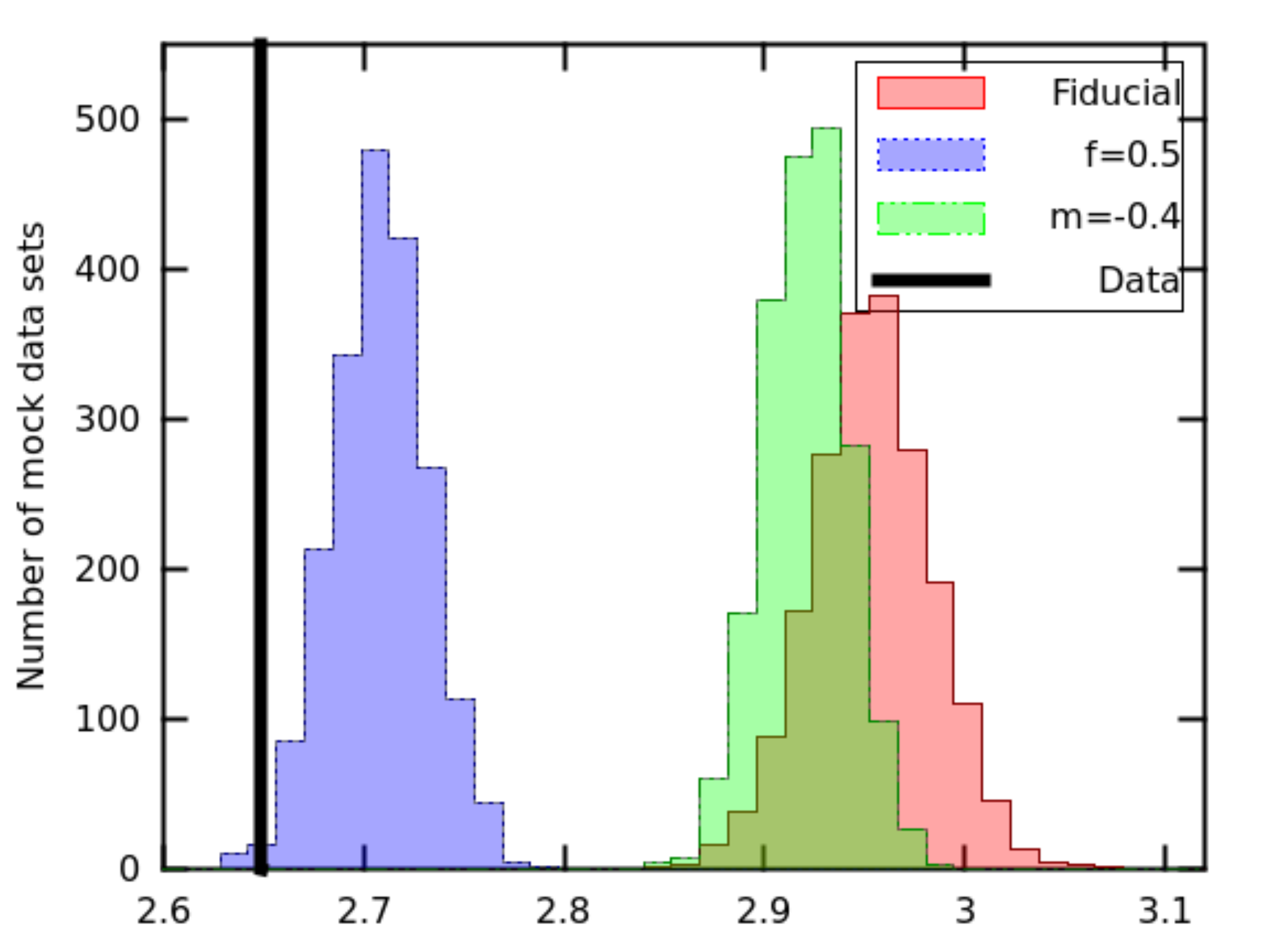}
    \label{fig:fig6a}
  }
  \subfigure[$\sigma(\textgoth{a})$]
  {
    \includegraphics[width=0.48\textwidth]{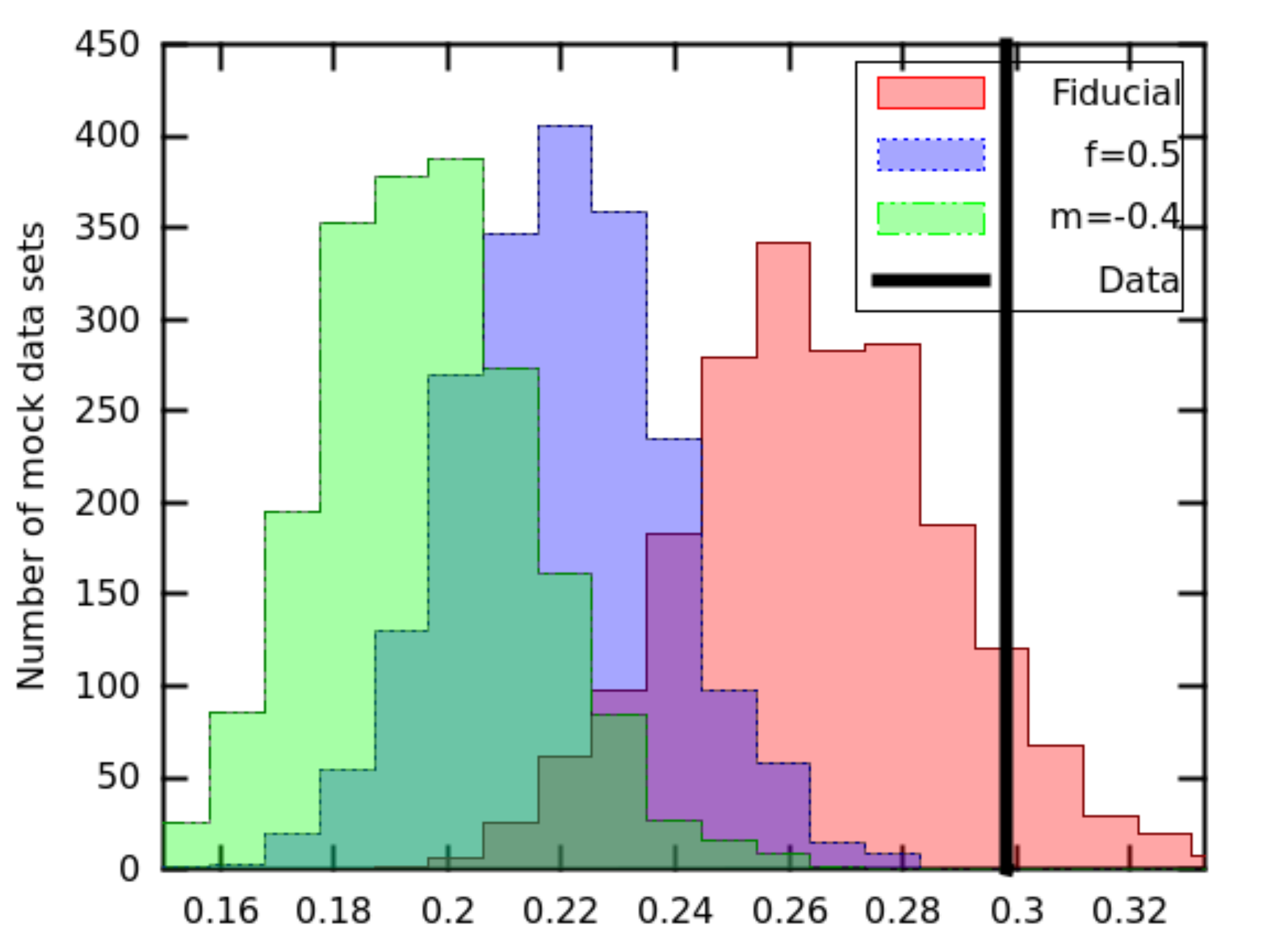}
    \label{fig:fig6b}
  }
  \subfigure[$\langle a \rangle_{2-2.5}$]
  {
    \includegraphics[width=0.48\textwidth]{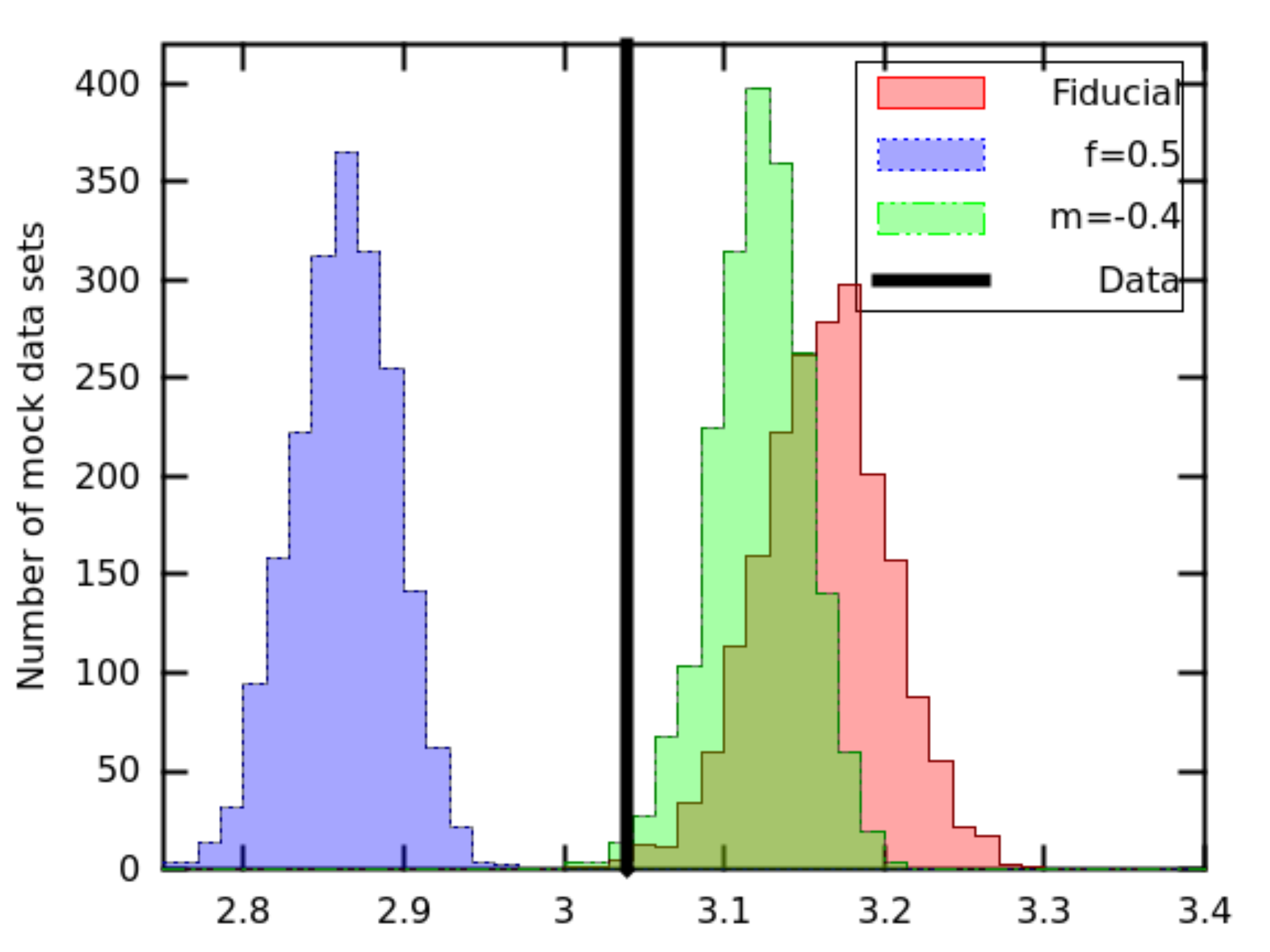}
    \label{fig:fig6c}
  }
  \subfigure[$\sigma(a)_{2-2.5}$]
  {
    \includegraphics[width=0.48\textwidth]{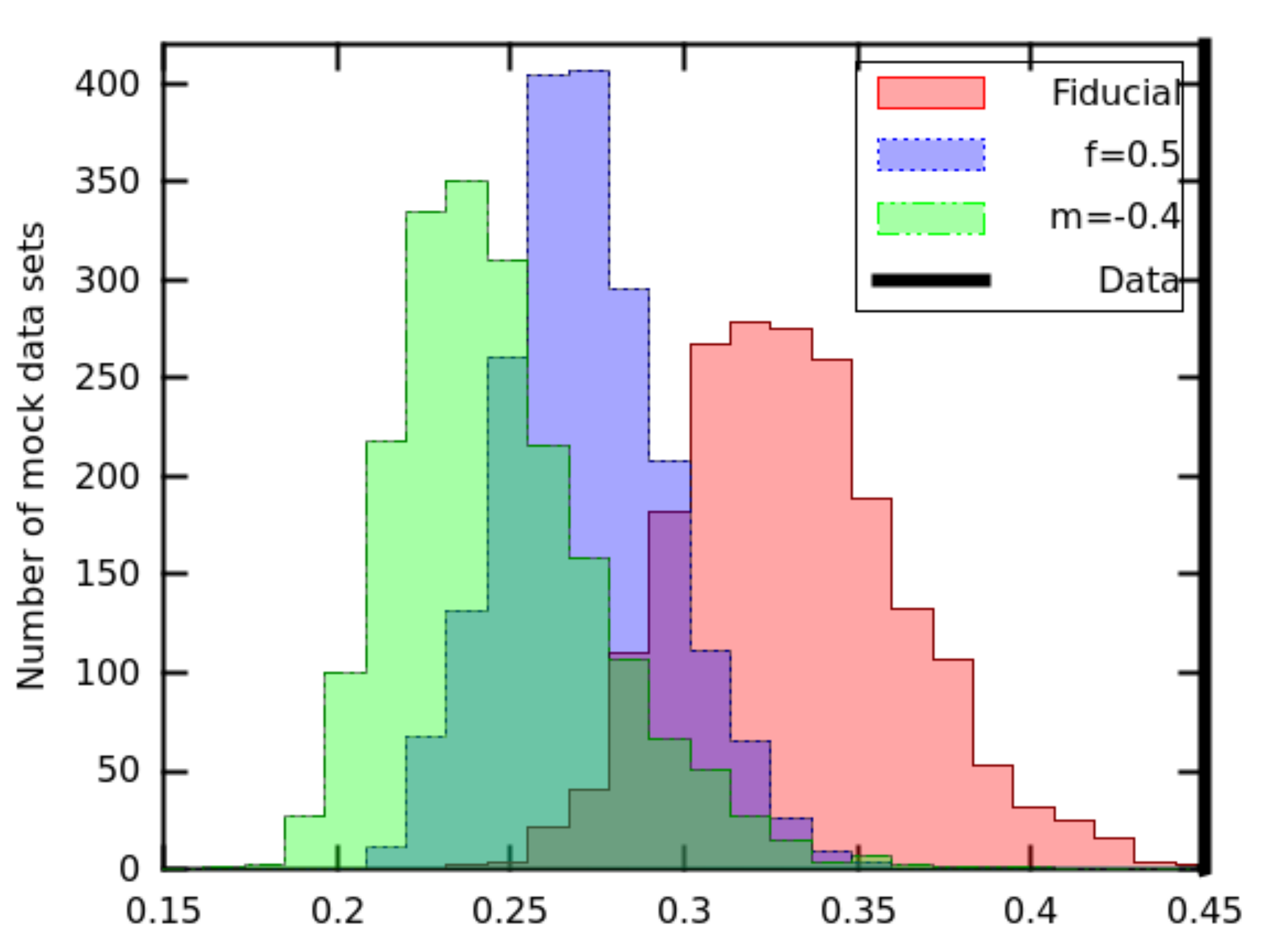}
    \label{fig:fig6d}
  }
  \caption{The distributions of four statistics quantifying the ``characteristic acceleration'' in the mock data (colours as in Fig.~\ref{fig:fig3}), compared to the corresponding values in the real data. Figs.~\ref{fig:fig6a} and~\ref{fig:fig6b} show the mean and standard deviation of $\textgoth{a}$, and Figs.~\ref{fig:fig6c} and~\ref{fig:fig6d} show the mean and standard deviation of $a_{2-2.5}$ (see Section~\ref{sec:comparison}). While the mean value of the ``characteristic acceleration'' may be matched within our parameter space, its spread is consistently underestimated.}
  \label{fig:fig6}
\end{figure*}

\begin{figure*}
  \subfigure[$\langle \Delta(\mathcal{D})_{M_*} \rangle$]
  {
    \includegraphics[width=0.44\textwidth]{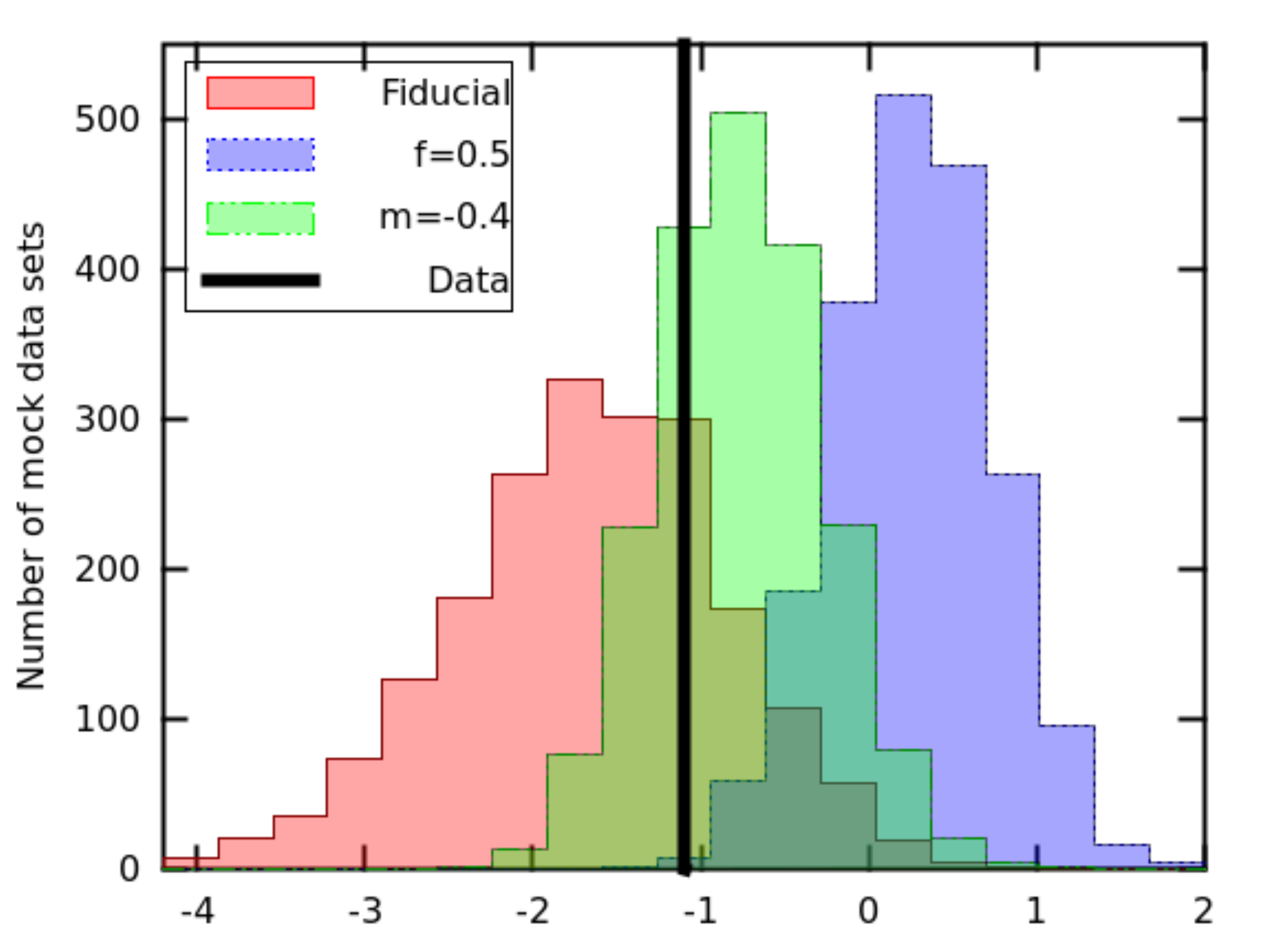}
    \label{fig:fig7a}
  }
  \subfigure[$\langle \Delta(\mathcal{D})_{\Delta R_\mathrm{d}} \rangle$]
  {
    \includegraphics[width=0.44\textwidth]{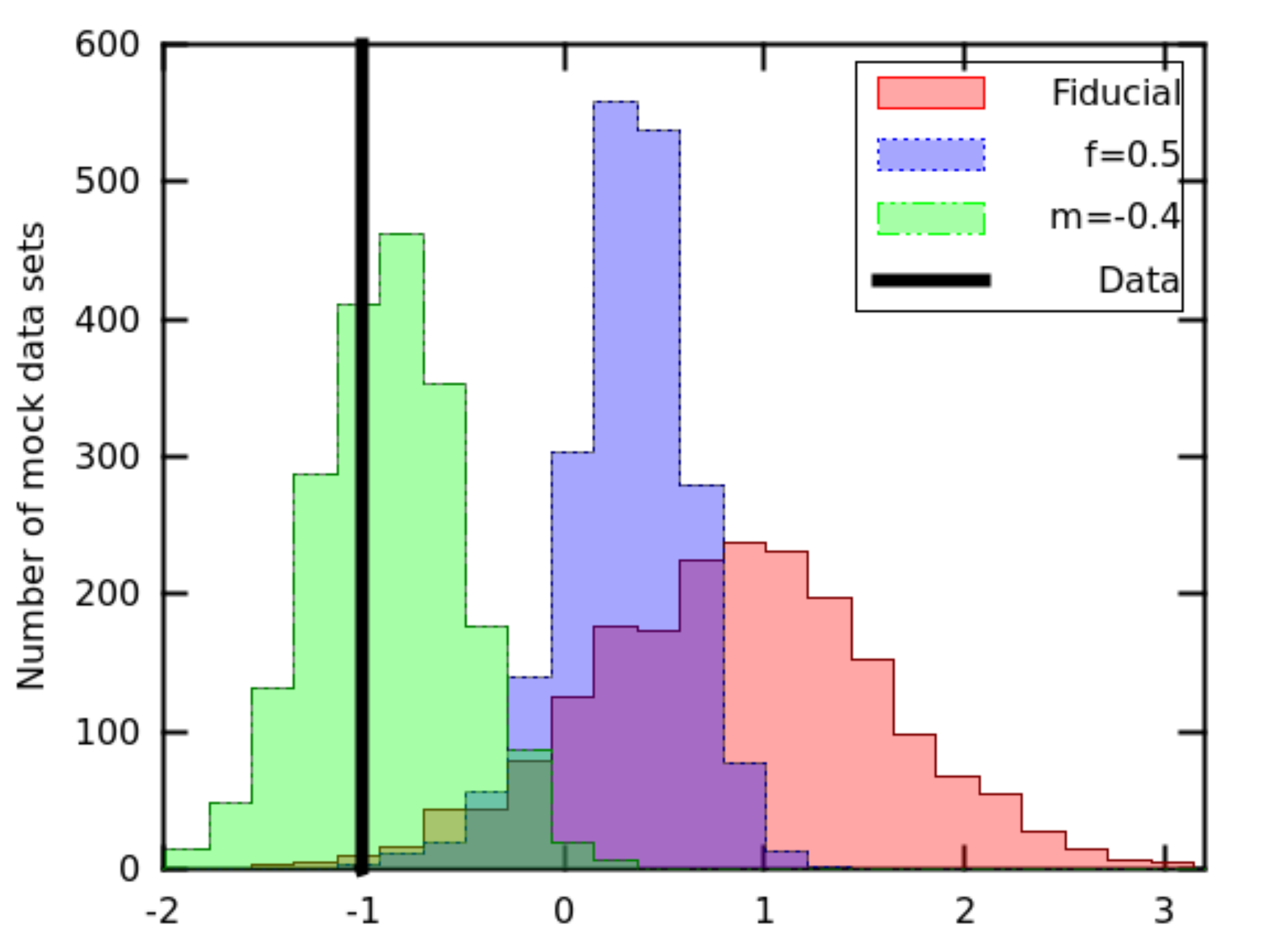}
    \label{fig:fig7b}
  }
  \subfigure[$\langle \Delta(\mathcal{D})_{\Delta M_\mathrm{gas}} \rangle$]
  {
    \includegraphics[width=0.44\textwidth]{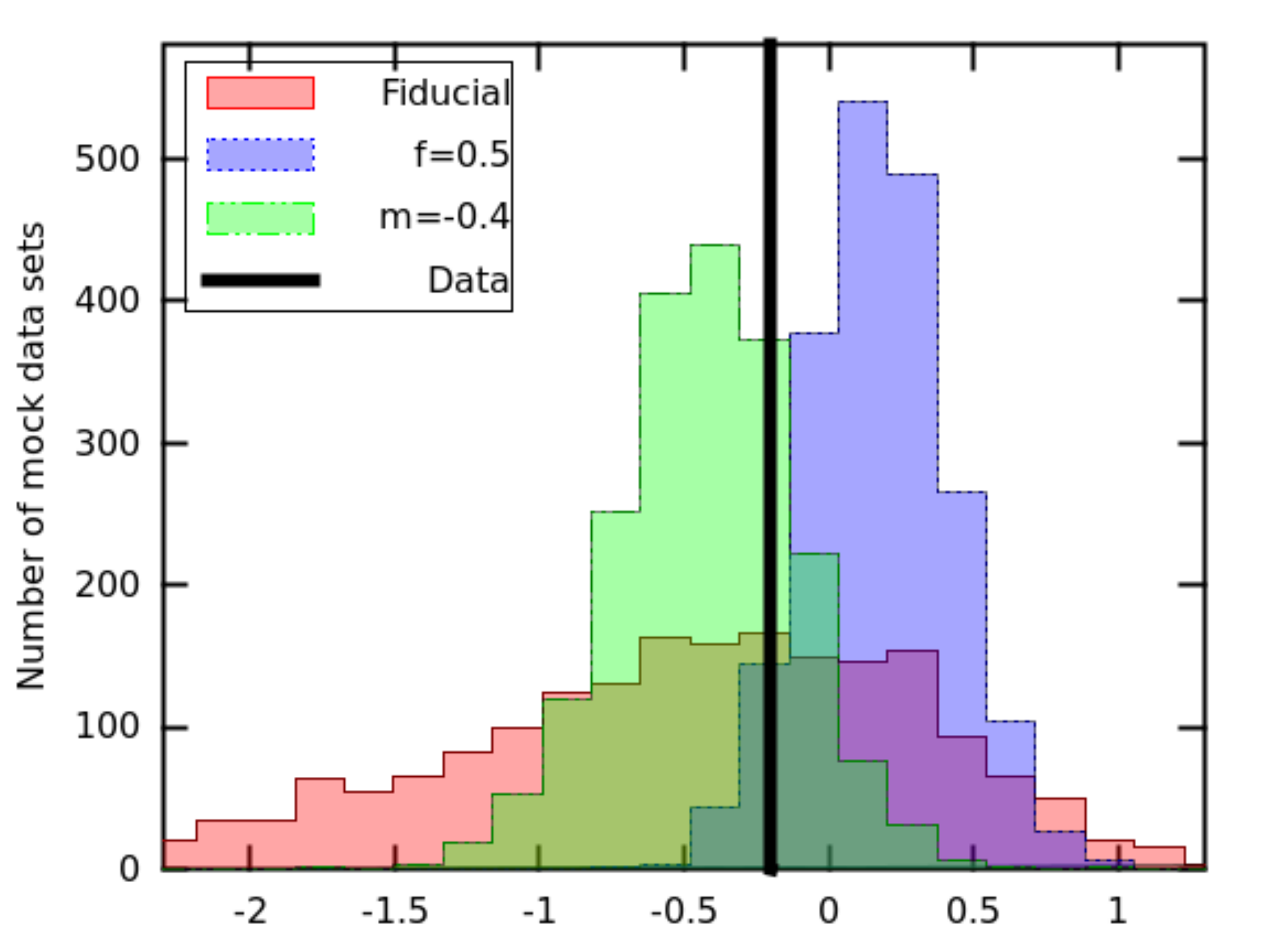}
    \label{fig:fig7c}
  }
  \subfigure[$\langle \Delta(\mathcal{D})_T \rangle$]
  {
    \includegraphics[width=0.44\textwidth]{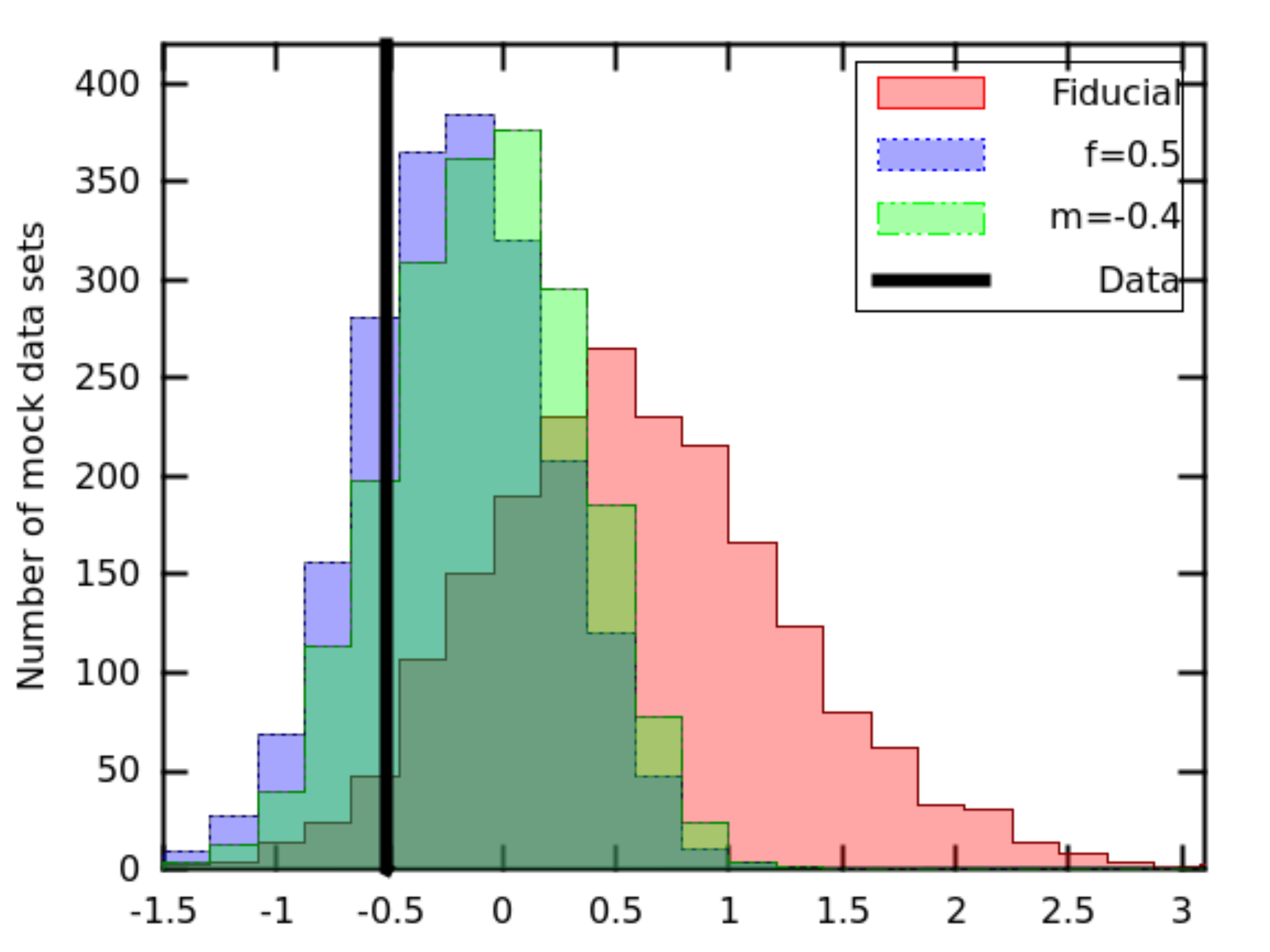}
    \label{fig:fig7d}
  }
  \subfigure[$\langle \Delta(\mathcal{D})_r \rangle$]
  {
    \includegraphics[width=0.44\textwidth]{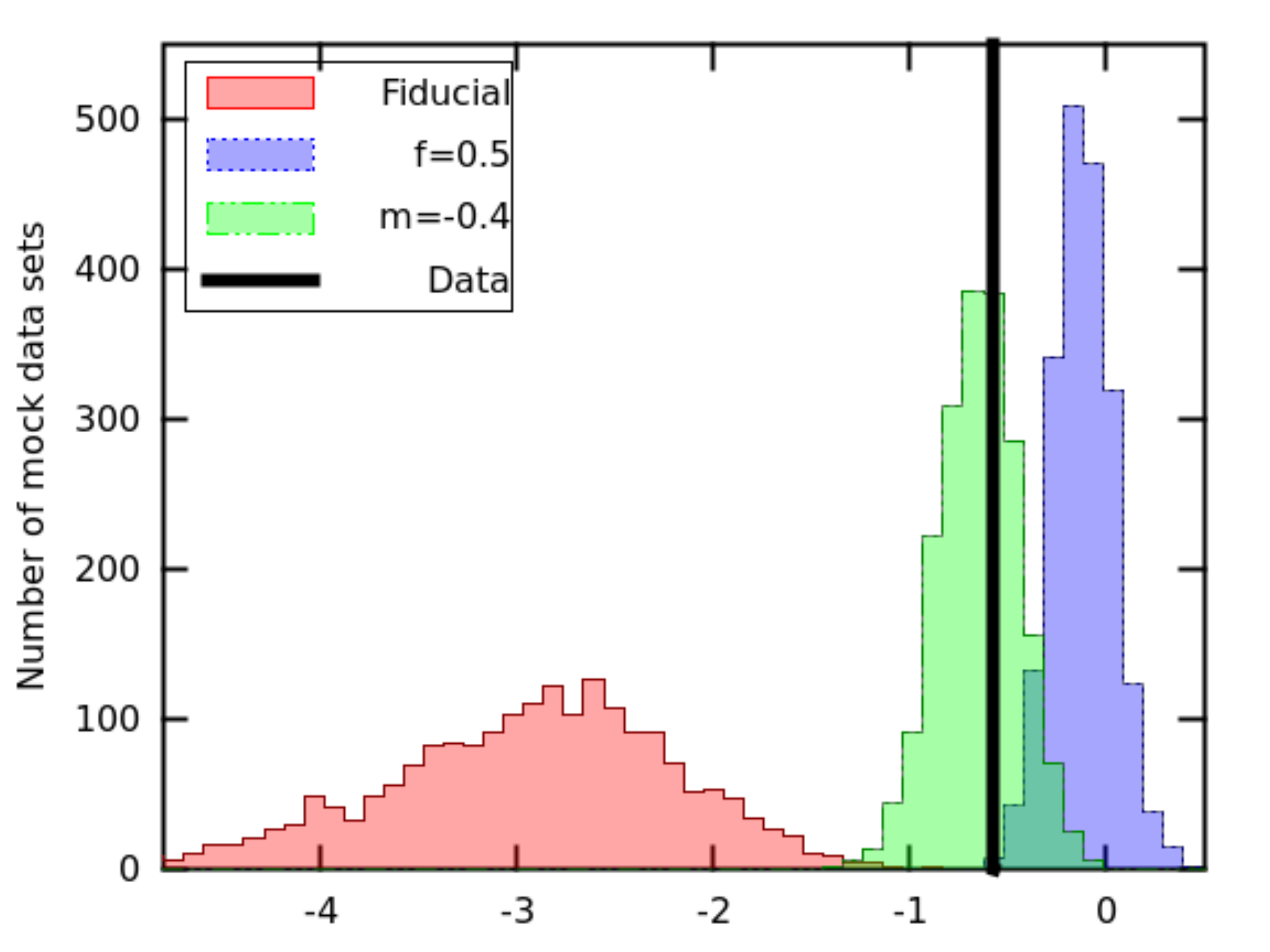}
    \label{fig:fig7e}
  }
  \caption{The weighted average difference in $\mathcal{D}$ at fixed $a$ ($\langle \Delta(\mathcal{D})_X \rangle$) between points with high and low values of various dynamically relevant variables $X$: $M_*$, $\Delta R_\mathrm{d}$, $\Delta M_\mathrm{gas}$, Hubble type $T$, and galactocentric radius $r$. The histograms show the distributions for 2000 mock data sets drawn from the corresponding models (colours as in Fig.~\ref{fig:fig3}), and the black line shows the value in the real data. While $\langle \Delta(\mathcal{D})_{M_*} \rangle$, $\langle \Delta(\mathcal{D})_{\Delta M_\mathrm{gas}} \rangle$ and $\langle \Delta(\mathcal{D})_T \rangle$ are matched well by a fiducial model, $\langle \Delta(\mathcal{D})_{\Delta R_\mathrm{d}} \rangle$ provides evidence for a correlation between galaxy size and halo mass or concentration at fixed $M_*$, and $\langle \Delta(\mathcal{D})_r \rangle$ for either this or strong concentration selection.}
  \label{fig:fig7}
\end{figure*}

We begin with a fiducial model in which $\alpha$ and AM scatter are chosen in accordance with the results of~\citealt{Lehmann} (0.6 and 0.16, respectively), and $\nu$, $f$, and $m$ are set to 0. This is the simplest AM expectation for the galaxy--halo connection, taking clustering measurements into account and assuming no coupling between baryons and dark matter or dependence of galaxy size or type on halo properties. The results are shown in Tables~\ref{tab:table1}--\ref{tab:table3} (second row) and Figs.~\ref{fig:fig2}--\ref{fig:fig7} (in red where models are compared).

The columns in Tables~\ref{tab:table1}--\ref{tab:table3} correspond to the statistics described in Section~\ref{sec:comparison}; those pertaining to MDAR shape and scatter are shown in Table~\ref{tab:table1}, characteristic acceleration in Table~\ref{tab:table2}, and correlation of MDAR residuals in Tables~\ref{tab:table3}. The first row lists the values in the \textsc{sparc} data, while subsequent rows contain the results for various models. In these cases, we give the modal average value of the statistic over all the Monte Carlo mock data sets, as well the minimal range enclosing 95 per cent of the results. If the value in the data lies outside these limits (and assuming negligible systematic error), then the null hypothesis that the data were drawn from the model may be rejected at the 95 per cent confidence level, according to that statistic.

Figure~\ref{fig:fig2} compares the real data in blue to a stack of 2000 Monte Carlo realisations of the theoretical galaxy population in magenta and a randomly chosen example mock data set in green. Fig.~\ref{fig:fig3a} shows the average mass discrepancy in each bin of acceleration, $\langle \mathcal{D} \rangle_i$, averaged over all mock data sets (line) with the associated $1\sigma$ scatter between mock data sets (shaded band); Fig.~\ref{fig:fig3b} is the analogue for the standard deviations $\sigma(\mathcal{D})_i$. Fig~\ref{fig:fig4} compares the weighted average standard deviation in $\mathcal{D}$, $\sigma_\mathrm{tot}$, for the mock and real data. Fig.~\ref{fig:fig5} is a normalised histogram of $\textgoth{a}$ for the real galaxies (red), and the mock galaxies stacked over all Monte Carlo realisations (blue). The insets show the distributions of $\mathcal{F}_+$ and $\mathcal{F}_-$ over the mock data sets, compared to the corresponding values for the real data. Figs.~\ref{fig:fig6a},~\ref{fig:fig6b},~\ref{fig:fig6c}, and~\ref{fig:fig6d} plot the distributions of $\langle \textgoth{a} \rangle$, $\sigma(\textgoth{a})$, $\langle a \rangle_{2-2.5}$, and $\sigma(a)_{2-2.5}$, respectively. Finally, Fig.~\ref{fig:fig7} shows the average offset $\langle \Delta(\mathcal{D})_X \rangle$ in $\mathcal{D}$ between subsets of the data split by $M_*$, $\Delta R_\mathrm{d}$, $\Delta M_\mathrm{gas}$, $T$, and $r$.

\afterpage{

\begin{table*}
  \begin{center}
    \begin{tabular}{l|c|c|c|c|c|c|c|c|c|c|c}
      \hline
      Model&($\alpha$, AM scatter, $f$, $m$, $\nu$) &$\langle \mathcal{D} \rangle_{1.5-1.8}$ &$\langle \mathcal{D} \rangle_{4.7-5}$ &$\sigma(\mathcal{D})_{1.5-1.8}$ &$\sigma(\mathcal{D})_{4.7-5}$ &$\sigma_\mathrm{tot}$\\ 
      \hline
\rule{0pt}{3.5ex}
      \textit{\fauxsc{sparc} data}&---		& $\mathit{10.1}$	& $\mathit{1.38}$	&$\mathit{5.5}$	& $\mathit{1.63}$	& $\mathit{1.63}$\\
\rule{0pt}{3.5ex}
      Fiducial&0.6, 0.16, 0, 0, 0			& \boldmath{$24.2^{+10.8}_{-7.4}$}	& $1.06^{+0.04}_{-0.05}$ &	\boldmath{$21.1^{+26.1}_{-11.4}$}	& $0.13^{+0.04}_{-0.02}$	& \boldmath{$5.15^{+3.47}_{-2.08}$}\\
\rule{0pt}{3.5ex}
      Moderate halo expansion&0.6, 0.16, 0, 0, $-0.5$	& \boldmath{$22.6^{+11.8}_{-7.3}$}	& $1.02^{+0.04}_{-0.04}$ &	\boldmath{$19.4^{+32.2}_{-12.6}$}	& $0.13^{+0.03}_{-0.02}$	& \boldmath{$4.30^{+4.33}_{-1.51}$}\\
\rule{0pt}{3.5ex}
      Strong halo expansion&0.6, 0.16, 0, 0, $-1$	    & \boldmath{$20.9^{+11.6}_{-6.9}$}	& $1.01^{+0.03}_{-0.04}$ &	\boldmath{$18.8^{+29.1}_{-13.2}$}	& $0.13^{+0.03}_{-0.03}$	& \boldmath{$4.40^{+4.24}_{-1.73}$}\\
\rule{0pt}{3.5ex}
      Adiabatic contraction&0.6, 0.16, 0, 0, 1			& \boldmath{$30.0^{+9.5}_{-9.0}$}	& \boldmath{$2.04^{+0.16}_{-0.25}$} &	\boldmath{$26.4^{+25.3}_{-13.4}$}	& $1.53^{+0.65}_{-0.65}$	& \boldmath{$5.59^{+3.94}_{-2.16}$}\\
\rule{0pt}{3.5ex}
      No AM scatter&0.6, 0, 0, 0, $-0.5$		& \boldmath{$22.4^{+13.0}_{-6.8}$}	& $1.01^{+0.05}_{-0.03}$ &	\boldmath{$18.3^{+33.1}_{-10.6}$}	& $0.13^{+0.03}_{-0.03}$	& \boldmath{$4.14^{+4.37}_{-1.33}$}\\
\rule{0pt}{3.5ex}
      Strong concentration selection&0.6, 0.16, 0.5, 0, $-0.5$		& $14.1^{+8.5}_{-3.6}$	& $1.02^{+0.04}_{-0.04}$ &	$6.0^{+11.2}_{-2.5}$	& $0.13^{+0.03}_{-0.03}$	& $1.65^{+1.03}_{-0.33}$\\
\rule{0pt}{3.5ex}
      Distinct haloes only&0.6, 0.16, 0, 0, $-0.5$			& \boldmath{$18.7^{+6.6}_{-4.4}$}	& $1.01^{+0.05}_{-0.04}$ &	\boldmath{$11.8^{+13.2}_{-5.8}$}	& $0.13^{+0.03}_{-0.03}$	& \boldmath{$2.96^{+1.88}_{-0.94}$}\\
\rule{0pt}{3.5ex}
      $\Delta R_\mathrm{d}-\Delta c$ anticorrelation&0.6, 0.16, 0, $-0.4$, $-0.5$	& \boldmath{$20.5^{+4.1}_{-4.1}$}	& $1.02^{+0.04}_{-0.04}$ &	\boldmath{$8.6^{+5.3}_{-2.7}$}	& $0.13^{+0.03}_{-0.03}$	& \boldmath{$2.26^{+0.67}_{-0.32}$}\\
      \hline
    \end{tabular}
  \caption{Statistics pertaining to the shape and scatter of the MDAR in the data (first row), and under various model assumptions (remaining rows). The numbers quoted for the models are modal averages across 2000 Monte Carlo realisations, along with the minimal bounds enclosing 95 per cent of the realisations ($2\sigma$). Bold indicates a prediction that we argue demonstrates a problem with the corresponding model, or motivates an alternative. See also Figs.~\ref{fig:fig3} and~\ref{fig:fig4}.}
  \label{tab:table1}
  \end{center}
\end{table*}

\begin{table*}
  \begin{center}
    \begin{tabular}{l|c|c|c|c|c|c|c|c|c|c|c|c|c}
      \hline
Model&($\alpha$, AM scatter, $f$, $m$, $\nu$) &$\langle \textgoth{a} \rangle$ &$\sigma(\textgoth{a})$ & $\mathcal{F}_+$ &$\mathcal{F}_-$ &$\langle a \rangle_{2-2.5}$ &$\sigma(a)_{2-2.5}$\\ 
      \hline
\rule{0pt}{3.5ex}
      \textit{\fauxsc{sparc} data}&---		& $\mathit{2.65}$	& $\mathit{0.30}$	& $\mathit{44}$	& $\mathit{26}$    & $\mathit{3.04}$  & $\mathit{0.45}$\\
\rule{0pt}{3.5ex}
      Fiducial&0.6, 0.16, 0, 0, 0			& \boldmath{$2.94^{+0.07}_{-0.05}$} & $0.25^{+0.05}_{-0.04}$ & \boldmath{$74^{+5}_{-6}$} & \boldmath{$14^{+4}_{-5}$}   & \boldmath{$3.18^{+0.06}_{-0.10}$}    & $0.31^{+0.09}_{-0.04}$\\
\rule{0pt}{3.5ex}
      Moderate halo expansion&0.6, 0.16, 0, 0, $-0.5$	& \boldmath{$2.85^{+0.06}_{-0.06}$} & $0.26^{+0.05}_{-0.06}$ & \boldmath{$67^{+6}_{-7}$} & \boldmath{$18^{+5}_{-5}$}   & $3.04^{+0.09}_{-0.09}$    & $0.31^{+0.07}_{-0.05}$\\
\rule{0pt}{3.5ex}
      Strong halo expansion&0.6, 0.16, 0, 0, $-1$			& \boldmath{$2.77^{+0.07}_{-0.05}$} & $0.25^{+0.05}_{-0.04}$ & \boldmath{$61^{+6}_{-7}$} & $21^{+7}_{-5}$   & $2.93^{+0.07}_{-0.08}$    & $0.32^{+0.06}_{-0.05}$\\
\rule{0pt}{3.5ex}
      Adiabatic contraction&0.6, 0.16, 0, 0, 1			& \boldmath{$3.22^{+0.09}_{-0.08}$} & $0.27^{+0.08}_{-0.07}$ & \boldmath{$102^{+5}_{-5}$} & \boldmath{$5^{+4}_{-2}$}   & \boldmath{$3.61^{+0.09}_{-0.11}$}    & $0.43^{+0.10}_{-0.05}$\\
\rule{0pt}{3.5ex}
      No AM scatter&0.6, 0, 0, 0, $-0.5$		& \boldmath{$2.86^{+0.06}_{-0.05}$} & $0.27^{+0.03}_{-0.05}$ & \boldmath{$67^{+5}_{-7}$} & \boldmath{$18^{+4}_{-6}$}   & $3.04^{+0.08}_{-0.07}$    & $0.33^{+0.06}_{-0.06}$\\
\rule{0pt}{3.5ex}
      Strong concentration selection&0.6, 0.16, 0.5, 0, $-0.5$	& $2.70^{+0.05}_{-0.04}$ & $0.22^{+0.03}_{-0.04}$ & \boldmath{$53^{+6}_{-4}$} & $27^{+6}_{-6}$   & \boldmath{$2.86^{+0.06}_{-0.06}$}    & $0.26^{+0.06}_{-0.04}$\\
\rule{0pt}{3.5ex}
      Distinct haloes only&0.6, 0.16, 0, 0, $-0.5$	& \boldmath{$2.82^{+0.06}_{-0.04}$} & $0.25^{+0.04}_{-0.04}$ & \boldmath{$63^{+5}_{-6}$} & \boldmath{$17^{+4}_{-5}$}   & $3.00^{+0.10}_{-0.05}$    & $0.28^{+0.05}_{-0.05}$\\
\rule{0pt}{3.5ex}
      $\Delta R_\mathrm{d}-\Delta c$ anticorrelation&0.6, 0.16, 0, $-0.4$, $-0.5$	& \boldmath{$2.91^{+0.04}_{-0.04}$} & $0.19^{+0.04}_{-0.04}$ & \boldmath{$71^{+4}_{-5}$} & \boldmath{$12^{+4}_{-3}$}   & $3.12^{+0.06}_{-0.06}$    & $0.23^{+0.08}_{-0.03}$\\
      \hline
    \end{tabular}
  \caption{Statistics pertaining to the ``characteristic acceleration'' of the MDAR in the data (first row), and under various model assumptions (remaining rows). Details as in Table~\ref{tab:table1}. See also Figs.~\ref{fig:fig5} and~\ref{fig:fig6}.}
  \label{tab:table2}
  \end{center}
\end{table*}

\begin{table*}
  \begin{center}
    \begin{tabular}{l|c|c|c|c|c|c|c|c|c|c|c}
      \hline
Model&($\alpha$, AM scatter, $f$, $m$, $\nu$) & $\langle \Delta(\mathcal{D})_{M_*} \rangle$ &$\langle \Delta(\mathcal{D})_{\Delta R_\mathrm{d}} \rangle$ &$\langle \Delta(\mathcal{D})_{\Delta M_\mathrm{gas}} \rangle$ &$\langle \Delta(\mathcal{D})_T \rangle$ &$\langle \Delta(\mathcal{D})_r \rangle$\\ 
      \hline
\rule{0pt}{3.5ex}
      \textit{\fauxsc{sparc} data}&---		& $\mathit{-1.09}$	& $\mathit{-1.01}$	& $\mathit{-0.19}$	& $\mathit{-0.51}$ & $\mathit{-0.57}$\\
\rule{0pt}{3.5ex}
      Fiducial&0.6, 0.16, 0, 0, 0			& $-1.32^{+1.15}_{-2.11}$	& \boldmath{$1.04^{+1.24}_{-1.79}$}	& $-0.29^{+1.30}_{-2.25}$	& $0.42^{+1.48}_{-1.12}$ & \boldmath{$-2.63^{+1.04}_{-1.89}$}\\
\rule{0pt}{3.5ex}
      Moderate halo expansion&0.6, 0.16, 0, 0, $-0.5$	& $-1.31^{+1.27}_{-1.82}$	& \boldmath{$0.56^{+1.59}_{-1.24}$}	& $-0.27^{+1.24}_{-2.24}$	& $0.13^{+1.63}_{-0.96}$ & \boldmath{$-2.53^{+1.60}_{-1.34}$}\\
\rule{0pt}{3.5ex}
      Strong halo expansion&0.6, 0.16, 0, 0, $-1$			& $-0.98^{+1.06}_{-1.95}$	& \boldmath{$0.61^{+1.57}_{-1.35}$}	& $-0.09^{+1.23}_{-2.35}$	& $0.10^{+1.42}_{-1.24}$ & \boldmath{$-2.02^{+1.33}_{-1.95}$}\\
\rule{0pt}{3.5ex}
      Adiabatic contraction&0.6, 0.16, 0, 0, 1			& $-2.26^{+1.51}_{-1.69}$	& \boldmath{$0.85^{+1.69}_{-1.32}$}	& $-0.72^{+1.59}_{-1.90}$	& \boldmath{$1.00^{+1.45}_{-1.26}$} & \boldmath{$-4.22^{+1.34}_{-1.61}$}\\
\rule{0pt}{3.5ex}
      No AM scatter&0.6, 0, 0, 0, $-0.5$		& $-1.06^{+0.94}_{-1.89}$	& \boldmath{$0.79^{+1.39}_{-1.39}$}	& $-0.61^{+1.74}_{-2.01}$	& $0.10^{+1.34}_{-1.01}$ & \boldmath{$-2.16^{+1.06}_{-1.95}$}\\
\rule{0pt}{3.5ex}
      Strong concentration selection&0.6, 0.16, 0.5, 0, $-0.5$			& $0.12^{+0.95}_{-1.01}$	& \boldmath{$0.27^{+0.56}_{-0.68}$}	& $-0.01^{+0.61}_{-0.36}$	& $-0.52^{+1.10}_{-0.60}$ & $-0.23^{+0.35}_{-0.26}$\\
\rule{0pt}{3.5ex}
      Distinct haloes only&0.6, 0.16, 0, 0, $-0.5$			& $-0.34^{+1.19}_{-1.26}$	& \boldmath{$0.26^{+1.33}_{-0.71}$}	& $0.17^{+0.71}_{-1.34}$	& $-0.36^{+1.19}_{-0.86}$ & $-0.91^{+0.72}_{-1.06}$\\
\rule{0pt}{3.5ex}
      $\Delta R_\mathrm{d}-\Delta c$ anticorrelation&0.6, 0.16, 0, $-0.4$, $-0.5$	& $-0.84^{+0.93}_{-0.99}$	& $-0.91^{+0.67}_{-0.75}$	& $-0.47^{+0.60}_{-0.64}$	& $0.04^{+0.58}_{-1.01}$ & $-0.67^{+0.37}_{-0.42}$\\
      \hline
    \end{tabular}
  \caption{Statistics pertaining to the dependence of the MDAR on various galaxy properties in the data (first row), and under various model assumptions (remaining rows). Details as in Table~\ref{tab:table1}. See also Fig.~\ref{fig:fig7}.}
  \label{tab:table3}
  \end{center}
\end{table*}
}

The results are as follows.

\begin{enumerate}

\item{} A visual inspection of Fig.~\ref{fig:fig2} reveals that the shape of the observed MDAR roughly traces the model expectation. $\mathcal{D}$ is low at high $a$ (where baryons dominate), and dark matter becomes increasingly important towards lower $a$. This result is known from the work of~\citet{vdB} and~\citet{DC}.

\item{} However, it is also evident from Fig.~\ref{fig:fig2} that both $\langle \mathcal{D} \rangle$ and $\sigma(\mathcal{D})$ are too high at low $a$; that is, there is an excess of dark matter in low-acceleration regions and an excessive spread between galaxies in the amount of dark matter. In a realisation of the model, many more points would be expected at $\mathcal{D} > 25$ than are observed. These results are confirmed in Fig.~\ref{fig:fig3}, and the third and fifth columns of Table~\ref{tab:table1}, where it is shown that the predicted $\mathcal{D}$ exceeds that observed for $a \lesssim 3.3$, making the relation appear on the whole too steep. The discrepancy is $3.8 \sigma$ in $\langle \mathcal{D} \rangle_{1.5-1.8}$, $2.7 \sigma$ in $\sigma(\mathcal{D})_{1.5-1.8}$, and $3.4\sigma$ in $\sigma_\mathrm{tot}$.

\item{} A transition occurs at $a \approx 3.3$, beyond which the simulated measurements become roughly coincident with the observations in $\langle \mathcal{D} \rangle$, and lie below them in $\sigma(\mathcal{D})$. This shows that standard AM (without halo contraction) readily accounts for the baryon domination of high-acceleration regions, and furthermore that variations in the dark matter content among galaxies at high $a$ are naturally small. In fact, it is seen from the fourth and sixth columns of Table~\ref{tab:table1} that the predicted $\langle \mathcal{D} \rangle$ and $\sigma(\mathcal{D})$ are both significantly too \emph{low} in the final bin, which implies at face value that high-acceleration regions ought to possess \emph{more} dark matter than allowed for by our fiducial model. However, the observed $\langle \mathcal{D} \rangle$ at high acceleration is likely consistent with one within the measurement uncertainty, which is quoted at $\lesssim 15$ per cent in $V_\mathrm{tot}$~\citep{SPARC} but receives additional contribution from uncertainties in disc inclination, mass-to-light ratio, distance, and three-dimensional baryon structure (private communication, F. Lelli). Including this would ameliorate the underprediction of $\langle \mathcal{D} \rangle_{4.7-5}$ and $\sigma(\mathcal{D})_{4.7-5}$ (as well as $\sigma(\textgoth{a})$ and $\sigma(a)_{2-2.5}$, below), but exacerbate the overprediction of $\sigma(\mathcal{D})_{1.5-1.8}$ and $\sigma_\mathrm{tot}$.

\item{} When the line traced by a galaxy in the $a-\mathcal{D}$ plane crosses $\mathcal{D}=3$, it does so at an acceleration $\textgoth{a}$ in the range $2.5 \lesssim a \lesssim 3.5$. The distribution of $\textgoth{a}$ values over all galaxies in a mock data set is comparable to that of the real data, although centred on a somewhat higher value (Figs.~\ref{fig:fig5a} and~\ref{fig:fig6a}, and column 3 of Table~\ref{tab:table2}). Remarkably, the spread in $\textgoth{a}$, $\sigma(\textgoth{a})$, is actually \emph{lower} in the model than in the data (Fig.~\ref{fig:fig6b} and column 4 of Table~\ref{tab:table2}), indicating a \emph{greater} degree of regularity in the acceleration beyond which dark matter becomes dynamically irrelevant. These results imply that certain aspects of the ``characteristic acceleration'' that galaxies are said to exhibit (such as a tight transition region between baryon and dark matter domination) are present also in AM mocks.

\item{} However, it can be seen from the inset of Fig.~\ref{fig:fig5a} and columns 5 and 6 of Table~\ref{tab:table2} that the number of galaxies with $\mathcal{D}>3$ everywhere ($\mathcal{F}_+$) is much larger in the mock data than the observations, and the number with $\mathcal{D}<3$ everywhere ($\mathcal{F}_-$) much smaller. This indicates a problem with the relative number of galaxies predicted to be fully baryon or dark matter dominated. We note however that using a larger measurement error in $\mathcal{D}$ would scatter more points below the $\mathcal{D}=3$ threshold, and hence reduce the predicted $\mathcal{F}_+$.

\item{} The other measure of characteristic acceleration yields similar results: $\langle a \rangle_{2-2.5}$ is overpredicted (though not as severely as $\langle \textgoth{a} \rangle$), and $\sigma(a)_{2-2.5}$ underpredicted (Figs.~\ref{fig:fig6c} and~\ref{fig:fig6d}, and columns 7 and 8 of Table~\ref{tab:table2}). This further supports the conclusion that the uniformity among galaxies in the acceleration at which dark matter becomes insignificant is not problematic. The differences between statistics (vi) to (vii) and (x) to (xi), however, illustrate the care that must be taken in defining precisely what one means by ``characteristic acceleration'' before its compatibility with model expectations can be assessed.

\item{} The observed values of $\langle \Delta(\mathcal{D})_{M_*} \rangle$, $\langle \Delta(\mathcal{D})_{\Delta M_\mathrm{gas}} \rangle$ and $\langle \Delta(\mathcal{D})_T \rangle$ are compatible with the corresponding theoretical distributions (Figs.~\ref{fig:fig7a},~\ref{fig:fig7c} and~\ref{fig:fig7d} and third, fifth and sixth columns of Table~\ref{tab:table3}). This implies that the relation of stellar to halo mass generated by AM is adequate, and that the MDAR does not favour a modification to stellar mass-based AM to include gas.\footnote{Note that an independence of halo properties on type, size or gas mass at fixed stellar mass (as in the fiducial model) does not imply $\langle \Delta(\mathcal{D})_X \rangle = 0$. This statistic depends on the contingent details of the baryonic mass distributions and sampling radii of the \textsc{sparc} galaxies. The only way to acquire information concerning these aspects of the galaxy--halo connection, therefore, is to compare mock and real data sets in which these distributions and radii are identical.} $\langle \Delta(\mathcal{D})_T \rangle$ provides weak evidence ($1.7\sigma$) for the earlier-type galaxies of the \textsc{sparc} sample occupying more massive or more concentrated haloes than allowed by the fiducial model. This offset is in the direction of indications from weak lensing and satellite kinematics~\citep{Puebla, Wojtak, Mandelbaum}. We caution however that these results are sensitive to the free parameters of the model, an issue to which we will return in Section~\ref{sec:varying_params}.

\item{} $\langle \Delta(\mathcal{D})_{\Delta R_\mathrm{d}} \rangle$ is typically larger in the model than the data (Fig.~\ref{fig:fig7b} and fourth column of Table~\ref{tab:table3}), and the observed value lies $2.3\sigma$ from the centre of the mock data distribution. In other words, the model predicts an offset between the MDARs of high and low surface brightness galaxies that is not observed. In Section~\ref{sec:varying_params}, we will remedy this by means of an anticorrelation between galaxy size and halo concentration at fixed stellar mass.

\item{} $\langle \Delta(\mathcal{D})_r \rangle$ is smaller in the mock data than the real data by $4.0\sigma$ (Fig.~\ref{fig:fig7e} and seventh column of Table~\ref{tab:table3}). This discrepancy -- a relative of the well-known cusp/core problem for dwarf galaxies -- indicates too much dark matter in the inner regions of the model galaxies.

\item{} The magnitudes of $\langle \Delta(\mathcal{D})_X \rangle$ in the data are all small ($\lesssim 1$). The average quoted measurement uncertainty of 7 per cent in $V_\mathrm{tot}$ corresponds to a 14 per cent uncertainty in $\mathcal{D}$, which equates to $\Delta(\mathcal{D}) \approx 0.5$ when averaged over the entire MDAR (and the \emph{total} measurement uncertainty is likely much larger). Hence the dependence of $\mathcal{D}$ on each $X$ is small enough to be fully accounted for by measurement error, and $\mathcal{D}$ is consistent with being a function of $a$ alone.

\end{enumerate}

\subsection{Variations to model parameters}
\label{sec:varying_params}

\subsubsection{Baryonic effects on dark matter distribution}

We now investigate the effect of varying the model parameter values, with a particular eye to mitigating the discrepancies noted above. We consider first the effect of modifications to the halo density profile due to disc formation, as parametrized by $\nu$. The excessive value of $\langle \mathcal{D} \rangle_{1.5-1.8}$ in the fiducial model, in addition to $\langle \Delta(\mathcal{D})_r \rangle$ being too low, suggests that halo expansion may be preferred. We therefore show in rows 3 and 4 of Tables~\ref{tab:table1}--\ref{tab:table3} the results of changing $\nu$ to $-0.5$ (favoured by previous studies of galaxy dynamics;~\citealt{Dutton_13, DW15, DW16}) and $-1$ (roughly the lowest value deemed plausible by~\citealt{DW15}), respectively. The lower $\nu$, the more dark matter is expelled from the baryon-dominated inner regions of galaxies.

Although $\langle \mathcal{D} \rangle_{1.5-1.8}$, $\sigma(\mathcal{D})_{1.5-1.8}$, $\sigma_\mathrm{tot}$, $\langle \textgoth{a} \rangle$, $\mathcal{F}_+$, $\mathcal{F}_-$, $\langle a \rangle_{2-2.5}$ and $\langle \Delta(\mathcal{D})_r \rangle$ all move closer to their observed values when $\nu$ is reduced, there remain significant discrepancies. This suggests -- at least within our parametrization -- that feedback is unlikely to fully reconcile the predicted and observed MDARs. For reference, we show in the fifth row of Tables~\ref{tab:table1}--\ref{tab:table3} the results of a model with standard adiabatic contraction ($\nu=1$). In accordance with many studies in the literature (e.g.~\citealt{D07, McGaugh_2007, DW15, DW16}), this clearly generates an excess of dynamical mass within the regions probed by kinematic measurements. In addition, the sensitivity of the amount of dark matter in high-acceleration regions to the baryon mass distribution significantly increases $\sigma(\mathcal{D})_{4.7-5}$, which for all other models simply reflects the measurement uncertainty. 

To facilitate comparison with literature studies of AM, our baseline for variations to other model parameters in the remainder of this section will be the ``moderate halo expansion'' model with $\nu=-0.5$.

\subsubsection{Scatter and proxy of abundance matching}

We explore next the possibility of reducing the predicted MDAR scatter by decreasing the AM scatter; the results of a model with AM scatter = 0 are shown in the sixth row of Tables~\ref{tab:table1}--\ref{tab:table3}. This has only a small effect on the statistics characterising the shape and scatter of the relation, indicating that $\sigma(\mathcal{D})$ originates almost entirely from the spread in $\mathcal{D}$ values generated by haloes of fixed proxy (due to scatter in $M_\mathrm{vir}$ and $c$ at fixed $v_\alpha$), rather than the spread in proxy values at fixed $M_*$. That $\sigma(\mathcal{D})_{1.5-1.8}$ and $\sigma_\mathrm{tot}$ are significantly overpredicted even without any scatter in the galaxy--halo connection demonstrates the extraordinary tightness of the MDAR. Since AM scatter = 0 is in any case ruled out by clustering~\citep{Lehmann}, we will not consider this model further. As in~\citet{DW16}, we find that the effect of varying $\alpha$ within the limits set by clustering measurements is small and does not significantly impact any of our statistics. We therefore conclude that varying the AM parameters cannot substantially improve the predicted MDAR.

\subsubsection{Morphology selection effects}

The scatter in the theoretical MDAR may be more efficiently reduced by imposing a correlation between galaxy morphology and halo concentration, effectively situating the late-type \textsc{sparc} galaxies in haloes less-concentrated than average. By skewing the dark matter profile to larger $r$, this may also be expected to mitigate the discrepancy in $\langle \Delta(\mathcal{D})_{r} \rangle$. As described in Section~\ref{sec:model}, this may be achieved by removing at each stellar mass some fraction $f$ of the highest concentration haloes before assigning haloes to the mock galaxies. Fig.~\ref{fig:fig5c} and the blue lines and histograms in Figs.~\ref{fig:fig3},~\ref{fig:fig4},~\ref{fig:fig6}, and~\ref{fig:fig7} show the results of removing the 50 per cent of highest concentration haloes at each stellar mass ($f=0.5$; see also row seven of Tables~\ref{tab:table1}--\ref{tab:table3}). The predictions of this model are approximately consistent with the observations in terms of $\langle \mathcal{D} \rangle_{1.5-1.8}$, $\sigma(\mathcal{D})_{1.5-1.8}$, $\sigma_\mathrm{tot}$, $\langle \textgoth{a} \rangle$, $\mathcal{F}_-$ and $\langle \Delta(\mathcal{D})_{r} \rangle$, indicating significant improvement over the fiducial case. We discuss this model further in Section~\ref{sec:problems}.

As an alternative selection model, we report in row 8 of Tables~\ref{tab:table1}--\ref{tab:table3} the case in which each \textsc{sparc} galaxy is required to occupy a distinct halo that is not located inside the virial radius of a larger halo. We find a moderate reduction in $\langle \mathcal{D} \rangle$ and $\sigma(\mathcal{D})$ across the acceleration range, and increase in $\langle \Delta(\mathcal{D})_{r} \rangle$, due to subhaloes being more concentrated than distinct haloes on average. In terms of its effect on the MDAR, this is a middle ground between the ``moderate halo expansion'' model and the $f=0.5$ model described above.

\subsubsection{Correlation of disc scale length with halo concentration}

None of the model alterations examined so far have had a significant impact on $\langle \Delta(\mathcal{D})_{\Delta R_\mathrm{d}} \rangle$ (Table~\ref{tab:table3}, column 4), which has remained in all cases at least $2-3 \sigma$ too high. This indicates that larger galaxies at fixed stellar mass ought to reside in less-massive haloes than have been assigned by the model, i.e. that $R_\mathrm{d}$ as well as $M_*$ ought to correlate with halo $c$ and/or $M_\mathrm{vir}$. We anticipated this possibility with the size--concentration correlation model described in Section~\ref{sec:model}. We now consider the case $m<0$, and tune $m$ until the discrepancy in $\langle \Delta(\mathcal{D})_{\Delta R_\mathrm{d}} \rangle$ is removed. We find $-0.7 \lesssim m \lesssim -0.1$ for the observed value of $\langle \Delta(\mathcal{D})_{\Delta R_\mathrm{d}} \rangle$ to lie within the 95 per cent confidence interval of the theoretical prediction, and show the best-fitting case $m = -0.4$ in Fig.~\ref{fig:fig5d} and Figs.~\ref{fig:fig3},~\ref{fig:fig4},~\ref{fig:fig6}, and~\ref{fig:fig7} (green), and Tables~\ref{tab:table1}--\ref{tab:table3} (row nine). This result shows that there is sufficient information in the MDAR to determine the correlations of galaxy size with halo variables. We discuss the ramifications of this result in Section~\ref{sec:constraints}.

An anticorrelation of $R_\mathrm{d}$ and $c$ has two further consequences. The first is a reduction in $\sigma(\mathcal{D})_{1.5-1.8}$ and $\sigma_\mathrm{tot}$, since a galaxy of given size can no longer occupy a halo with any concentration. These remain greater than the observed values, however, even when the correlation between $c$ and $\Delta R_\mathrm{d}$ is made very tight. This suggests that the spread in $M_\mathrm{vir}$ alone at fixed $M_*$, $R_\mathrm{d}$ and $c$ overpredicts the MDAR scatter. The second consequence is an increase in $\langle \Delta (\mathcal{D})_r \rangle$, bringing it into good agreement with the data. Since large galaxies now inhabit less-concentrated haloes, their $\mathcal{D}$ values are increased at large $r$ and decreased at small $r$. This outweighs the opposite effect for small galaxies that tend not to have measurements at large $r$ where $\mathcal{D}$ would be reduced.

\section{Discussion}
\label{sec:discussion}

\subsection{Comparison with the literature}
\label{sec:literature}

There have been few theoretical studies of the MDAR, and none to the level of detail presented here. Those that do discuss the relation use it to argue either that $\Lambda$CDM-based models must not be able to account for the detailed dynamical properties of galaxies~\citep{Famaey_McGaugh, Kroupa_Falsification, TTP, Kroupa}, or, in reaction to such claims, that the relation emerges readily from basic assumptions concerning galaxy formation~\citep{vdB, DC}. By means of a detailed analysis of the statistical properties of the relation and its quantitative agreement with specific assumptions for the galaxy--halo connection, we intend our work to overcome this dichotomy and achieve a non-binary appreciation of the relation's significance. In this section, we discuss the MDAR literature in light of our results. A recurring theme will be the assertion that a given model ``can fit the MDAR'' or ``cannot in principle fit the MDAR,'' where in fact the associated analysis warrants only the conclusion that it can or cannot fit one particular MDAR statistic -- and even that not established quantitatively.

A series of papers~\citep{MG98, McGaugh_MDAcc, Famaey_McGaugh, ThirdLaw, TTP} argues that the MDAR poses a serious fine-tuning problem for all $\Lambda$CDM-based galaxy formation models. In particular, it is claimed that neither the tightness of the observed MDAR nor the bottoming out of $\mathcal{D}$ at high $a$ can be accounted for. We confirm that the small scatter of the MDAR is a serious problem, although in our analysis the discrepancy is not as severe as suggested for example by fig.~3 of~\citet{TTP}. This is because~\citet{TTP} considers as a priori plausible correlations between galaxy and halo variables that violate the tenets of AM, while we take these tenets for granted.  While the improvement of our models over those of~\citet{TTP} may therefore be considered a success of AM (and halo expansion), it simultaneously underscores the need to understand the physical processes responsible for generating the tight galaxy--halo connection it assumes.

Two further successes of our models over those above are the prediction of low $\langle \mathcal{D} \rangle$ at high acceleration, and of a tight distribution of $a$ values at which $\mathcal{D}$ becomes small. These results derive from the combination of the parameter values of concordance cosmology, the stellar-to-halo mass fractions of AM, halo expansion, and realistic galaxy sizes and baryonic mass profiles. Thus while an AM model for the galaxy--halo connection is not uniquely favoured by these aspects of the MDAR, it does at least offer one successful approach. Indeed, that a ``characteristic acceleration'' of dark matter is of order $c \: H_0$ -- regardless of the details of the galaxy--halo connection -- is perhaps not so surprising if we consider that the average dark matter density is an $\mathcal{O}(1)$ multiple of $\rho_\mathrm{crit}$, which is proportional to $H_0^2$.

Nevertheless, many of the claims of~\citet{Famaey_McGaugh},~\citet{ThirdLaw}, and~\citet{TTP} are borne out by our investigation. Despite a clear prediction of baryon domination at high $a$, our framework fails to reproduce several facets of the ``acceleration scale'' that marks the transition to this regime. None of our models simultaneously match the mean acceleration at $2 < \mathcal{D} < 2.5$ and the mean acceleration at which individual galaxies cross $\mathcal{D}=3$, and models without strong concentration selection produce the wrong relative frequency of galaxies with $\mathcal{D}>3$ and $\mathcal{D}<3$ across their rotation curves. In addition, $\sigma(\mathcal{D})_{1.5-1.8}$ and $\sigma_\mathrm{tot}$ are naturally overpredicted, and remain so under highly conservative assumptions about galaxy formation, including no scatter in the galaxy--halo connection and a significant reduction in the spread of halo concentrations associated with galaxies of given $M_*$ and $R_\mathrm{d}$. We verify also that $\mathcal{D}$ is indeed compatible with being a function of $a$ alone ($|\langle \Delta(\mathcal{D})_X \rangle| \lesssim 1$).

\citet{vdB} are concerned in part to show that galaxy formation scenarios with correct phenomenology in other respects must fit the MDAR. They present a model with feedback parameters tuned to match the Tully--Fisher relation, among others, and use it to generate 40 mock galaxies that they sample at regular intervals to produce a theoretical MDAR. They find that $\mathcal{D}$ is a tighter function of $a$ than $r$ or $V$, and that it tends to 1 at high $a$. They claim on the basis of these results that the model successfully exhibits a characteristic acceleration scale. Although our study qualitatively agrees with several results of~\citet{vdB}, we believe their claim of success to be too strong -- such a conclusion could only be warranted by a statistical analysis in which a mock data set with all the important features of the observed MDAR is not extraordinary given the parent galaxy--halo population. In particular, claiming success for \emph{any} reasonable model precludes using the relation to constrain models, which we believe we have shown to be both possible and fruitful.

\citet{Kroupa} examine the MDAR of galaxies produced in several CDM and WDM hydrodynamical simulations, and conclude that the predicted and observed relations are incompatible. This conclusion is supported by an application of the Wilcoxon signed rank test that quantifies the agreement of the ranks of $\mathcal{D}$ in the observed and simulated data at given $a$. Although this test clearly demonstrates that the expected MDAR is on the whole normalised too high (a conclusion corroborated here), we caution that it need not be considered as decisive as the authors claim: since $\mathcal{D}$ is reduced to its rank, a simulated MDAR consistently lying only slightly above the observations will appear as incompatible with them as one lying much higher, despite the fact that uncertainties in either data set may suffice to make them fully compatible according to any goodness-of-fit test that uses $\mathcal{D}$ directly. Furthermore, the failure of individual simulations does not imply impossibility within the simulational framework. We suspect in addition that~\citet{Kroupa} use MOND mass-to-light ratios that have no justification in $\Lambda$CDM. Nevertheless, we concur that the scatter in the expected MDAR is on the whole too high, and for the reasons given: there is a large spread in halo properties for given baryonic properties. We have placed this discrepancy on firm numerical footing, and furthermore brought to light the conditions required for a quantitative solution. Whether or not such conditions may be reasonably achieved by hydrodynamical simulations in $\Lambda$CDM remains to be seen.

\citet{DC} construct a semi-empirical model for the MDAR using average galaxy scaling relations from observations and halo scaling relations from $N$-body simulations. Although similar in spirit, our methodology improves upon that of~\citet{DC} in several respects: we use an AM model that matches clustering measurements, we consider selection effects and correlations of halo variables with galaxy properties besides stellar mass, we give mock galaxies $M_\mathrm{gas}$, $R_\mathrm{d}$, $R_\mathrm{d,gas}$ and $V^2_\mathrm{bar}(r)$ values identical on a galaxy-by-galaxy basis with the data and sample their rotation curves at the same radii, we take the correlations between halo variables directly from a state-of-the-art $N$-body simulation rather than approximating them by power-laws with fixed log-normal scatter, and finally we make quantitative comparisons of specific MDAR features.

Finally,~\citet{Santos-Santos} compare the observed MDAR to that of 22 galaxies generated by the MaGICC and CLUES hydrodynamical simulations. Although the shape and scatter of the relation cannot be quantified with such a small sample, both seem to be in approximate agreement with the observations. We believe that an empirical framework offering complete transparency concerning its constituent elements provides a better handle than hydrodynamical simulation on the aspects of the galaxy--halo connection responsible for important features of the MDAR. Nevertheless, simulations will produce correlations between galaxy and halo variables that lie outside our parametrizations, making a direct comparison with the MDAR valuable and complementary to our approach. We hope that future comparisons will utilise the statistics we have developed.

\subsection{Parameter constraints and model requirements}
\label{sec:constraints}

The MDAR imposes several interesting constraints on AM-based galaxy formation models in $\Lambda$CDM, and provides evidence concerning a number of important correlations of the galaxy--halo connection. We discuss these issues here.

First, in agreement with ~\citet{Borriello},~\citet{D07, Dutton_13},~\citet{DW15, DW16}, and~\citet{DC}, the observed $\mathcal{D}$ values fit more comfortably with expanded than contracted haloes: a high value of $\nu$ (such as that corresponding to standard adiabatic contraction) degrades the fit for the majority of our statistics. We find little sensitivity to the AM parameters within the ranges allowed by clustering measurements, although more generally the tightness of the MDAR argues for a small AM scatter. We find no evidence from $\langle \Delta(\mathcal{D})_{M_*} \rangle$ for deviations from the AM stellar mass--halo mass relation, or from $\langle \Delta(\mathcal{D})_{\Delta M_\mathrm{gas}} \rangle$ for preferring AM based on total baryonic mass to the conventional stellar mass-based scheme. $\langle \Delta(\mathcal{D})_T \rangle$ provides at best weak evidence for early-type galaxies occupying more massive or more concentrated haloes than late-types at fixed stellar mass; this result could be strengthened using a larger sample of galaxies with a greater range of morphology.

The small offset between the MDARs of galaxies with the same stellar mass but different size requires at the $\sim 2 \sigma$ level that larger galaxies be placed in less-massive or less-concentrated haloes at fixed $M_*$ than is allowed for by standard AM. We have constructed a simple toy model for such a relation, and found that the observed value of $\langle \Delta(\mathcal{D})_{\Delta R_\mathrm{d}} \rangle$ implies an average correlation $\Delta c \propto m \cdot \Delta R_\mathrm{d}$, with $-0.7 \lesssim m \lesssim -0.1$ (95 per cent confidence) and best-fitting value $m = -0.4$. The weakness of this correlation lends some justification to the assumption used in many semi-analytic and empirical models that galaxy size is uncorrelated with halo properties at fixed $M_*$ (e.g.~\citealt{D11, Dutton_13, DW16, DC}), and indeed models that make this assumption yield approximate agreement with observed $\Delta V - \Delta R$ correlations (e.g. the tilt of the Fundamental Plane in the case of~\citealt{Dutton_13} and~\citealt{DW16}) while those that impose strong $\Delta R_\mathrm{d}-\Delta c$ or $\Delta R_\mathrm{d}-\Delta M_\mathrm{vir}$ correlations a priori do not (e.g. the correlation of velocity and size residuals in~\citealt{DW15}). We stress in any case that a correlation of galaxy size with halo properties has important dynamical consequences and no strong prior (indeed, it is predicted by the celebrated~\citealt{MMW} model), and should therefore be explicitly considered in any galaxy formation model that seeks to explain galaxies' $M-V-R$ scaling relations. In an upcoming paper (Desmond et al., in preparation), we will show that the \textsc{eagle} hydrodynamical simulation~\citep{EAGLE} produces a $\Delta R_\mathrm{d}-\Delta c$ correlation in approximate agreement with the limits we set here.

Remarkably, the ``characteristic acceleration'' at which $\mathcal{D}$ becomes small is predicted to be no less uniform among galaxies than that observed for all the models we consider, and hence the sharpness of this feature is not constraining.

Predictions for many of our statistics ($\langle \mathcal{D} \rangle_{1.5-1.8}$, $\sigma(\mathcal{D})_{1.5-1.8}$, $\sigma_\mathrm{tot}$, $\langle \textgoth{a} \rangle$, $\mathcal{F}_+$, $\mathcal{F}_-$, and $\langle \Delta(\mathcal{D})_r \rangle$) are substantially improved by situating the \textsc{sparc} galaxies in a subset of the total halo population highly biased towards low concentration. We discuss the plausibility of this model further in the following section.

\subsection{Remaining problems}
\label{sec:problems}

We have identified several aspects of the MDAR that cannot be matched -- or can be matched only with extreme difficulty -- in our framework. Foremost among these is the scatter, which we find to be too high in the theory both at low $a$ ($\sigma(\mathcal{D})_{1.5-1.8}$) and in an averaged sense ($\sigma_\mathrm{tot}$), regardless of the details of the AM scheme, the halo response to disc formation, and the correlation between galaxy size and halo concentration at fixed stellar mass. This is a particularly pressing problem because AM imposes a tight galaxy--halo connection by construction, and there is no guarantee that a full galaxy formation theory would naturally produce a correlation between $M_*$ and $v_\alpha$ with only $\sim 0.16$ dex scatter. Indeed, the MDAR scatter is overpredicted even in the case of a perfectly monotonic galaxy--halo connection, when AM scatter is switched off. In addition, we have neglected observational uncertainties from disc inclination, mass-to-light ratio, distance, and three-dimensional baryon structure, which would increase the intrinsic tightness of the relation still further. Finally, we have been careful to ensure that there are no differences between the local baryonic properties of mock and real galaxies or the radii at which their rotation curves are sampled, both of which could artificially inflate the predicted scatter. Our conclusions concerning $\sigma(\mathcal{D})$ ought therefore to be highly conservative. The unexpected regularity of galaxy kinematics has been noted previously in the context of the baryonic Tully--Fisher relation, most recently for the \textsc{sparc} sample in~\citet*{BTFR}.

A possible solution to this problem is to suppose that the \textsc{sparc} galaxies occupy a highly biased subset of the full halo population. We find that excluding the $f \approx 50$ per cent of haloes with greatest concentration at each stellar mass reduces $\sigma(\mathcal{D})_{1.5-1.8}$ and $\sigma_\mathrm{tot}$ to acceptable levels, in addition to mitigating the offset of $\mathcal{D}$ at high and low $r$ as measured by $\langle \Delta(\mathcal{D})_r \rangle$. In line with~\citet{DW15}, it is tempting to interpret this as a correlation between morphology and host halo structure. That such a model may be cashed out in detail, however, is doubtful, for the following four reasons:

\begin{enumerate}

\item{} The \textsc{sparc} galaxies span the range from S0 to Im/BCD, a morphological subset that accounts for more than half of the total galaxy population in the range $10^{6.7} \: M_\odot < M_* < 10^{11.5} \: M_\odot$ at $z=0$ (e.g.~\citealt{Bamford, Henriques}).

\item{} Even supposing one could justify discarding 50 per cent of haloes, it is implausible that these should be the exact haloes with highest concentration. A non-monotonic selection function would require a further increase in $f$ to be effective.

\item{} When the fraction of haloes removed is constant over stellar mass, $\langle \mathcal{D} \rangle$ and $\sigma(\mathcal{D})$ are also reduced for $a \gtrsim 3.3$, where they already lie at or below the observed values in the fiducial model. Since the high-acceleration regime is mostly populated by high-mass galaxies, this would suggest that the fraction of haloes removed ought to be lower at higher $M_*$, yet the early-type fraction rises with $M_*$.

\item{} The approximate agreement between the observed and predicted $\langle \Delta(\mathcal{D})_T \rangle$ shows that the average differences in the masses and mass distributions of haloes of galaxies with different morphology in the \textsc{sparc} sample cannot be large. Forcing galaxies of earlier type to inhabit significantly more concentrated haloes would reduce $\langle \Delta(\mathcal{D})_T \rangle$, perhaps out of consistency with the data.

\end{enumerate}

The second most significant challenge is the excess $\langle \mathcal{D} \rangle$ predicted at low $a$. This continues a long line of results indicating that the amount of dark matter in galaxies predicted by galaxy formation models applied to $N$-body simulations is on the whole too high, a problem that manifests itself in many and varied forms (e.g.~\citealt{Klypin, Moore, Bournaud, Kuzio, TBTF, TBTF2, Karach, Weinberg}). We note that while this discrepancy can be remedied in the case of galaxy one-point scaling relations -- the Tully--Fisher and Faber--Jackson relations -- by means of a halo expansion parametrized in the same way as adiabatic contraction~\citep{DW15, DW16}, the radius-dependent information in the MDAR reveals this model to be inadequate in detail. No value of $\nu$ in the range that correctly predicts the Tully--Fisher normalisation, for example, can generate a value of $\langle \mathcal{D} \rangle_{1.5-1.8}$ as low as that observed.

Low-acceleration points come from low surface brightness galaxies (LSBs) and the outer regions of high surface brightness galaxies (HSBs). Two unique features of the \textsc{sparc} sample are its high proportion of LSBs and its wide range of stellar mass (down to $M_* = 10^{6.7} M_\odot$). Tully--Fisher and Faber--Jackson data sets, on the other hand, tend to be dominated by HSBs and extend no lower than $M_* \approx 10^9 M_\odot$. The excess mass discrepancy at low $a$ is largely due to low mass LSBs: these galaxies are placed in haloes that are too massive. There are hints that this would impact the baryonic Tully--Fisher relation were it extended to low mass~\citep{D12, Papastergis}, and the discrepancy receives full expression in the Too Big To Fail~\citep{TBTF,TBTF2} and cusp-core~\citep{cusp-core} problems at the dwarf scale. Potential solutions to those problems apply here also, most prominently the possibility of strong feedback. Indeed, the low value of $\langle \Delta(\mathcal{D})_r \rangle$ in the fiducial model (indicating an excess of dark matter at low radii relative to high) may also be evidence of feedback. To preferentially impact low-acceleration regions, a successful scheme would not only need to move dark matter outwards, but do so preferentially in low mass LSBs. If feedback is powered by star formation, such galaxies might be expected to have a smaller impact on their haloes, although some simulations (e.g.~\citealt{DC1}) do predict such an effect. We stress however that simply reducing $\mathcal{D}$ by moving dark matter outwards is not enough: as discussed above, the spread in $\mathcal{D}$ must also be reduced, for example by homogenising the haloes. It is unclear whether this is expected.

The discrepancy in $\langle \mathcal{D} \rangle_{1.5-1.8}$ may also be alleviated by modifying the galaxies' stellar masses. We have assumed that the stellar masses assigned by AM are identical to those of the \textsc{sparc} sample, although in practice they have been derived using different assumptions for galaxy photometry, mass-to-light ratios and IMF. Since the AM stellar-to-halo mass ratio peaks around $M_* = 10^{10.5} M_\odot$, if the galaxies at the low-mass end of the \textsc{sparc} sample in fact correspond to galaxies of higher stellar mass in our model then the predicted mass discrepancies would be reduced at low $a$. This effect is degenerate with a modification to AM that places lower mass galaxies in less-massive haloes. It is also possible that extrapolating the stellar and halo mass function below their resolution limits ($M_* \approx 10^9 M_\odot$) could introduce systematic errors at the low-$a$ end of the MDAR.

If none of these modifications prove effective, non-standard dark matter interactions may be required (e.g.~\citealt*{Blanchet, SIDM, Suarez, Khoury}).

Finally, we note that not all of our statistics describing the ``acceleration scale'' of the MDAR are simultaneously reproduced by any one of our models. This issue is to some extent degenerate with the overall shape of the predicted relation, which has too steep a logarithmic slope (Fig.~\ref{fig:fig3}), and shows that more work is required to fully understand this phenomenon.

\subsection{Implications for MOND}
\label{mond}

Since the MOND prediction for the MDAR has less freedom than that of $\Lambda$CDM, the relation offers less potential to constrain it; on the flip-side, this makes the MOND prediction easier to rule out. We note three ways in which our analysis has relevance for MOND.

\begin{itemize}

\item{} \textit{Prima facie}, $\langle \mathcal{D} \rangle > 1$ at high $a$ implies the existence of missing mass in the Newtonian regime, violating a basic principle of MOND. Since we have shown that this is equally unexpected for AM, however, it is likely attributable to measurement error. This may derive from uncertainties in disc inclination, mass-to-light ratio, distance, or the three-dimensional distributions of the baryons.

\item{} Absent systematic uncertainty, $\mathcal{D}$ depends \emph{only} on $a$ in MOND, and hence MDAR residuals may not correlate with any other variable. In our analysis, this implies $\langle \Delta(\mathcal{D})_X \rangle = 0$ for each of $X \in \{M_*, \Delta R_\mathrm{d}, \Delta M_\mathrm{gas}, T, r\}$. As noted in Section~\ref{sec:fiducial}, the magnitudes of each of these variables in the data are $\lesssim 1$ and hence compatible with 0 within the measurement uncertainty, supporting the hypothesis that $\mathcal{D}$ is a function of $a$ alone. This finding is made somewhat more significant by the fact that our model predictions for $|\langle \Delta(\mathcal{D})_X \rangle|$ significantly exceed 1 in some cases.

\item{} In a Bayesian model comparison sense, treating the MDAR as one set of observations contributing to the overall probability of a theory, an increase in the likelihood achievable by $\Lambda$CDM models disfavours MOND, and vice versa. We have shown that the high-acceleration behaviour of the relation cannot discriminate against AM mocks. On the other hand, we have considerably strengthened the argument that the relation's scatter is lower than expected in standard galaxy formation, especially at low acceleration.

\end{itemize}

\section{Conclusion}
\label{sec:conclusion}

The MDAR provides a map between the distribution of a galaxy's baryonic and dark matter, and therefore contains crucial information about the galaxy--halo connection. In this paper we have laid the groundwork for extracting this information for use in evaluating models of galaxy formation. We have analysed the MDAR using a set of 16 statistics that quantify its four most important features: its shape, its scatter, the presence of a ``characteristic acceleration scale'' beyond which mass discrepancy consistently goes to $\sim 1$, and the correlation of its residuals with other galaxy properties. In addition to using these statistics to focus discussion of the observed relation itself, we have engaged them to construct a data-driven framework for the galaxy--halo connection. Building upwards from the simplest case of stellar mass-based abundance matching in $\Lambda$CDM, we have successively incorporated selection effects, a correlation between galaxy size and halo concentration, and a mass-dependent prescription for the impact of disc formation on halo density profiles. Comparing to data from the \textsc{sparc} sample, our most significant findings are as follows.

\begin{itemize}

\item{} A basic AM model readily accounts for several features of the MDAR, including its approximate overall shape, its normalisation and scatter at high acceleration, and the independence of its residuals on stellar and gas mass.

\item{} Nevertheless, the predicted MDAR has significantly too high a normalisation and scatter at low acceleration, and too high a scatter in an averaged sense over the whole relation. This remains true under highly conservative assumptions for galaxy formation, including no scatter in the galaxy--halo connection and a significant reduction in the spread of halo concentrations associated with galaxies of given $M_*$ and $R_\mathrm{d}$. This indicates too much dark matter mass predicted in the outer regions of high surface brightness galaxies and (especially) in low surface brightness galaxies, and too large a spread therein. In addition, dark matter is more concentrated towards the centres of galaxies than the MDAR suggests. These discrepancies argue for halo expansion in response to disc formation, and a quantitative resolution requires also the exclusion of a large fraction ($\sim 50$ per cent) of the haloes with highest concentration at each stellar mass.

\item{} We devise six statistics to capture aspects of the ``characteristic acceleration scale'' the MDAR is sometimes said to exhibit, describing its acceleration behaviour -- and that of individual galaxies within it -- at low mass discrepancy. Although our models cannot simultaneously reproduce the observed values of each of these statistics, we find no grounds for the claim that the transition region between baryonic and dark matter domination is sharper than expected by standard galaxy formation in $\Lambda$CDM.

\item{} The MDAR may be used to detect correlations of halo properties with (at least) three galaxy properties at fixed stellar mass: disc size, Hubble type, and gas mass. Our analysis provides weak evidence for an anticorrelation of halo mass or concentration with galaxy size and type at fixed stellar mass ($\sim 2.3 \sigma$ and $\sim 1.7 \sigma$ respectively), but no evidence for such a correlation with gas mass.

\end{itemize}

We hope that this work will stimulate interest in the MDAR as a source of information about the galaxy--halo connection. Looking forward, we identify three ways in which further progress could be made. First, additional models and assumptions need to be tested against the relation to the level of rigour achieved here. It is unclear to what extent the outputs of many hydrodynamical simulations or semi-analytic models, for example, are consistent with MDAR statistics. Specific implementations of the correlations of halo properties with galaxy variables besides $M_*$ (e.g. AM using total baryonic mass, or the angular momentum partition model of~\citealt{MMW}) should be tested individually. Second, a firmer theoretical basis needs to be given for the empirical galaxy--halo correlations argued for by the MDAR; this may be possible by mapping hydrodynamical simulations onto phenomenological frameworks such as ours. Finally, an increase in the size and precision of MDAR data sets may be expected to yield a considerable gain in the constraining power of analyses of this type, pinning down the values of the statistics in the real data and reducing the widths of their distributions in the mock data. The full power of the MDAR likely remains to be harnessed.

\section*{Acknowledgements}

I thank Xavier Hernandez, Federico Lelli, Stacy McGaugh and Risa Wechsler for fruitful discussions during the course of this work, and Simon Foreman for comments on the manuscript. I am particularly grateful to Federico Lelli for access to and guidance with the \textsc{sparc} data. The paper benefited also from the suggestions from an anonymous referee.

This work made use of one of the Dark Sky simulations, which were produced using an INCITE 2014 allocation on the Oak Ridge Leadership Computing Facility at Oak Ridge National Laboratory. I thank the Dark Sky collaboration for creating and providing access to this simulation, and Sam Skillman and Yao-Yuan Mao for running \textsc{rockstar} and \textsc{consistent trees} on it, respectively. Additional computation was performed at SLAC National Accelerator Laboratory.

This work received partial support from the U.S.\ Department of Energy under contract no. DE-AC02-76SF00515.

\bsp	


\begin{thebibliography}{10}

\bibitem[Bamford et al.(2009)]{Bamford}
Bamford S.~P. et al., 2009, MNRAS, 393, 1324 

\bibitem[\protect\citeauthoryear{Behroozi, Conroy \& Wechsler}{Behroozi et al.}{2010}]{Behroozi_2010}
Behroozi P.~S., Conroy C., Wechsler R.~H., 2010, ApJ, 717, 379 

\bibitem[\protect\citeauthoryear{Behroozi, Wechsler \& Wu}{Behroozi et al.}{2013}]{Rockstar_1}
Behroozi P.~S., Wechsler R.~H., Wu H.-Y., 2013, ApJ, 762, 109 

\bibitem[\protect\citeauthoryear{Behroozi et al.}{2013}]{Rockstar_2}
Behroozi P.~S., Wechsler R.~H., Wu H.-Y., Busha M., Klypin A., Primack J., 2013, ApJ, 763, 18 

\bibitem[Bernardi et al.(2013)]{Bernardi_SMF}
Bernardi M., Meert A., Sheth R.~K., Vikram V., Huertas-Company M., Mei F., Shankar F., 2013, MNRAS, 436, 697

\bibitem[Blanchet \& Le Tiec(2008)]{Blanchet}
Blanchet L., Le Tiec A., 2008, Phys. Rev. D, 78, 024031 

\bibitem[\protect\citeauthoryear{Blumenthal et al.}{1986}]{Blumenthal}
Blumenthal G.~R., Faber S.~M., Flores R., Primack J.~R., 1986, ApJ, 301, 27

\bibitem[\protect\citeauthoryear{Borriello, Salucci \& Danese}{Borriello et al.}{2003}]{Borriello}
Borriello A., Salucci P., Danese L., 2003, MNRAS, 341, 1109 

\bibitem[Bournaud et al.(2007)]{Bournaud}
Bournaud F., 2007, Science, 316, 1166 

\bibitem[\protect\citeauthoryear{Boylan-Kolchin, Bullock \& Kaplinghat}{Boylan-Kolchin et al.}{2011}]{TBTF}
Boylan-Kolchin M., Bullock J.~S., Kaplinghat M., 2011, MNRAS, 415, L40

\bibitem[\protect\citeauthoryear{Boylan-Kolchin, Bullock \& Kaplinghat}{Boylan-Kolchin et al.}{2012}]{TBTF2}
Boylan-Kolchin M., Bullock J.~S., Kaplinghat M., 2012, MNRAS, 422, 1203 

\bibitem[\protect\citeauthoryear{Conroy, Wechsler \& Kravtsov}{Conroy et al.}{2006}]{Conroy}
Conroy C., Wechsler R.~H., Kravtsov A.~V., 2006, ApJ, 647, 201 

\bibitem[de Blok(2010)]{cusp-core}
de Blok W.~J.~G., 2010, Adv. Astron., 2010, 789293

\bibitem[Desmond(2012)]{D12}
Desmond H., 2012, preprint (arXiv:1204.1497)

\bibitem[Desmond \& Wechsler(2015)]{DW15}
Desmond H., Wechsler R.~H., 2015, MNRAS, 454, 322 

\bibitem[Desmond \& Wechsler(2016)]{DW16}
Desmond H., Wechsler R.~H., 2016, preprint (arXiv:1604.04670) 

\bibitem[Di Cintio et al.(2014)]{DC1}
Di Cintio A., Brook C.~B., Dutton A.~A., Macci{\`o} A.~V., Stinson G.~S., Knebe A., 2014, MNRAS, 441, 2986 

\bibitem[Di Cintio \& Lelli(2016)]{DC}
Di Cintio A., Lelli F., 2016, MNRAS, 456, L127 

\bibitem[Dutton et al.(2007)]{D07}
Dutton A.~A., van den Bosch F.~C., Dekel A., Courteau S., 2007, ApJ, 654, 27 

\bibitem[Dutton et al.(2011)]{D11}
Dutton A.~A. et al., 2011, MNRAS, 416, 322 

\bibitem[Dutton et al.(2013)]{Dutton_13}
Dutton A.~A., Macci{\`o} A.~V., Mendel J.~T., Simard L., 2013, MNRAS, 432, 2496 

\bibitem[Famaey \& McGaugh(2012)]{Famaey_McGaugh}
Famaey B., McGaugh S.~S., 2012, Living Rev. Relativ., 15, 10 

\bibitem[Gnedin et al.(2004)]{Gnedin_2004}
Gnedin O.~Y., Kravtsov A.~V., Klypin A.~A., Nagai D., 2004, ApJ, 616, 16
 
\bibitem[\protect\citeauthoryear{Gnedin et al.}{2011}]{Gnedin_2011}
Gnedin O.~Y., Ceverino D., Gnedin N.~Y., Klypin A.~A., Kravtsov A.~V., Levine R., Nagai D., Yepes G., 2011, preprint (arxiv:1108.5736)

\bibitem[Governato et al.(2010)]{Governato}
Governato F. et al., 2010, Nature, 463, 203 

\bibitem[Guo et al.(2010)]{Guo}
Guo Q., White S., Li C., Boylan-Kolchin M., 2010, MNRAS, 404, 1111 

\bibitem[Hearin \& Watson(2013)]{Hearin_Watson}
Hearin A.~P., Watson D.~F., 2013, MNRAS, 435, 1313 

\bibitem[Hearin et al.(2014)]{CAM}
Hearin A.~P., Watson D.~F., Becker M.~R., Reyes R., Berlind A.~A., Zentner A.~A., 2014, MNRAS, 444, 729 

\bibitem[Henriques et al.(2015)]{Henriques}
Henriques B.~M.~B., White S.~D.~M., Thomas P.~A., Angulo R., Guo Q., Lemson G., Springel V., Overzier R, 2015, MNRAS, 451, 2663 

\bibitem[Janz et al.(2016)]{Janz}
Janz J., Cappellari M., Romanowsky A.~J., Ciotti L., Alabi A., Forbes D., 2016, MNRAS, 461, 2367

\bibitem[Jennings et al.(2016)]{Jennings}
Jennings E., Wechsler R.~H., Skillman S.~W., Warren M.~S., 2016, MNRAS, 457, 1076 

\bibitem[Karachentsev(2012)]{Karach}
Karachentsev I.~D., 2012, Astrophys. Bull., 67, 123 

\bibitem[Berezhiani \& Khoury(2016)]{Khoury}
Berezhiani L., Khoury J., 2015, PRD, 92, 103510

\bibitem[Klypin et al.(1999)]{Klypin}
Klypin A., Kravtsov A.~V., Valenzuela O., Prada F., 1999, ApJ, 522, 82 

\bibitem[Kravtsov et al.(2004)]{Kravtsov}
Kravtsov A.~V., Berlind A.~A., Wechsler R.~H., Klypin A.~A., Gottlober S., Allgood B., Primack J.~R., 2004, ApJ, 609, 35 

\bibitem[Kroupa(2012)]{Kroupa_Falsification}
Kroupa P., 2012, Publ. Astron. Soc. Aust., 29, 395 

\bibitem[Kuzio de Naray \& Spekkens(2011)]{Kuzio}
Kuzio de Naray R., Spekkens K., 2011, ApJ, 741, L29 

\bibitem[Lehmann et al.(2015)]{Lehmann}
Lehmann B.~V., Mao Y.-Y., Becker M.~R., Skillman S.~W., Wechsler R.~H., 2015, preprint (arXiv:1510.05651)

\bibitem[\protect\citeauthoryear{Lelli, McGaugh \& Schombert}{Lelli et al.}{2016a}]{BTFR}
Lelli F., McGaugh S.~S., Schombert J.~M., 2016, ApJ, 816, L14 

\bibitem[\protect\citeauthoryear{Lelli, McGaugh \& Schombert}{Lelli et al.}{2016b}]{SPARC}
Lelli F., McGaugh S.~S., Schombert J.~M., 2016, preprint (arXiv:1606.09251)

\bibitem[Mandelbaum et al.(2016)]{Mandelbaum}
Mandelbaum R., Wang W., Zu Y., White S., Henriques B., More S., 2016, MNRAS, 457, 3200

\bibitem[Marin et al.(2011)]{ABC}
Marin J.-M., Pudlo P., Robert C.~P., Ryder R., 2011, preprint (arXiv:1101.0955) 

\bibitem[\protect\citeauthoryear{Mashchenko, Wadsley \& Couchman}{Mashchenko et al.}{2008}]{Mash}
Mashchenko S., Wadsley J., Couchman H.~M.~P., 2008, Science, 319, 174 

\bibitem[McGaugh \& de Blok(1998)]{MG98}
McGaugh S.~S., de Blok, W.~J.~G., 1998, ApJ, 499, 41 

\bibitem[McGaugh(1999)]{MG99}
McGaugh S.~S., 1999, Galaxy Dynamics -- A Rutgers Symposium, Astron. Soc. Pacific Conf. Ser., ed. D.R. Merritt, M. Valluri, J.A. Sellwood, Vol. 182 (San Francisco: ASP), p. 528

\bibitem[McGaugh(2004)]{McGaugh_MDAcc}
McGaugh S.~S., 2004, ApJ, 609, 652 

\bibitem[McGaugh et al.(2007)]{McGaugh_2007}
McGaugh S.~S., de Blok W.~J.~G., Schombert J.~M., Kuzio de Naray R., Kim J.~H., 2007, ApJ, 659, 149 

\bibitem[McGaugh(2014)]{ThirdLaw}
McGaugh S.~S., 2014, Galaxies, 2, 601 

\bibitem[McGaugh(2015)]{TTP}
McGaugh S.~S., 2015, Can. J. Phys., 93, 250 

\bibitem[Milgrom(1983a)]{Milgrom1}
Milgrom M., 1983, ApJ, 270, 365

\bibitem[Milgrom(1983b)]{Milgrom2}
Milgrom M., 1983, ApJ, 270, 371 

\bibitem[Milgrom(1983c)]{Milgrom3}
Milgrom M., 1983, ApJ, 270, 384 

\bibitem[\protect\citeauthoryear{Mo, Mao \& White}{Mo et al.}{1998}]{MMW}
Mo H.J., Mao S., White S.D.M., 1998, MNRAS, 295, 319

\bibitem[Moore et al.(1999)]{Moore}
Moore B., Quinn T., Governato F., Stadel J., Lake G., 1999, MNRAS, 310, 1147 

\bibitem[Moster et al.(2010)]{Moster}
Moster B.~P., Somerville R.~S., Maulbetsch C., van den Bosch F.~C., Macci{\`o} A.~V., Naab T., Oser L., 2010, ApJ, 710, 903 

\bibitem[\protect\citeauthoryear{Papastergis, Adams \& van der Hulst}{Papastergis et al.}{2016}]{Papastergis}
Papastergis E., Adams E.~A.~K., van der Hulst J.~M., 2016, A\&A, 593, A39

\bibitem[Pontzen \& Governato(2012)]{Pontzen}
Pontzen A., Governato F., 2012, MNRAS, 421, 3464 

\bibitem[Reddick et al.(2013)]{Reddick}
Reddick R.~M., Wechsler R.~H., Tinker J.~L., Behroozi P.~S., 2013, ApJ, 771, 30

\bibitem[Rocha et al.(2013)]{SIDM}
Rocha M., Peter A.~H.~G., Bullock J.~S., Kaplinghat M., Garrison-Kimmel S., Onorbe J., Moustakas L., 2013, MNRAS, 430, 81 

\bibitem[Rodr{\'{\i}}guez-Puebla et al.(2011)]{Puebla} 
Rodr{\'{\i}}guez-Puebla A., Avila-Reese V., Firmani C., Col{\'{\i}}n P., 2011, Rev. Mex. Astron. Astrofis., 47, 235 

\bibitem[Sanders(1990)]{Sanders_MDAcc}
Sanders R.~H., 1990, A\&AR, 2, 1

\bibitem[Santos-Santos et al.(2016)]{Santos-Santos}
Santos-Santos I.~M., Brook C.~B., Stinson G., Di Cintio A., Wadsley J., Dominquez-Tenreiro R., Gottlober S., 2016, MNRAS, 455, 476 

\bibitem[Schaye et al.(2015)]{EAGLE}
Schaye J. et al., 2015, MNRAS, 446, 521

\bibitem[Skillman et al.(2014)]{DarkSky}
Skillman S.~W., Warren M.~S., Turk M.~J., Wechsler R.~H., Holz D.~E., Sutter P.~M., 2014, preprint (arXiv:1407.2600)

\bibitem[\protect\citeauthoryear{Su{\'a}rez, Robles \& Matos}{Su{\'a}rez et al.}{2014}]{Suarez}
Su{\'a}rez A., Robles V.~H., Matos T., 2014, Astrophys. Space Sci. Proc., Vol. 38, p. 107

\bibitem[Tiret \& Combes(2009)]{Tiret_Combes}
Tiret O., Combes F., 2009, A\&A, 496, 659

\bibitem[van den Bosch \& Dalcanton(2000)]{vdB}
van den Bosch F.~C., Dalcanton J.~J., 2000, ApJ, 534, 146

\bibitem[Warren(2013)]{Warren13}
Warren M.~S., 2013, Proc. Int. Conf. High Perform. Comput. Netw. Storage Anal., (New York: ACM), p. 72

\bibitem[Weinberg et al.(2013)]{Weinberg}
Weinberg D.~H., Bullock J.~S., Governato F., Kuzio de Naray R., Peter A.~H.~G., 2013, PNAS 112, 40

\bibitem[Wojtak \& Mamon(2013)]{Wojtak}
Wojtak R., Mamon G.~A., 2013, MNRAS, 428, 2407

\bibitem[Wu \& Kroupa(2015)]{Kroupa}
Wu X., Kroupa P., 2015, MNRAS, 446, 330

\end{thebibliography}
\end{document}